\documentclass[10pt]{article}

\usepackage{amsmath,amssymb,graphicx}

\newcommand{\om}{\omega}
\newcommand{\ph}{\varphi}
\newcommand{\ga}{\gamma}
\newcommand{\ct}{c^*}
\newcommand{\st}{s_0^*}
\newcommand{\sob}{\overline{s}_0}
\newcommand{\qs}{q^{\mathrm{s}}}
\newcommand{\qm}{q^{\mathrm{M}}}
\newcommand{\cmone}{c_1^{\mathrm{M}}}
\newcommand{\cmtwo}{c_2^{\mathrm{M}}}
\newcommand{\pa}{\partial}
\newcommand{\lb}{\overline{\ell}}
\newcommand{\cb}{\overline{c}}
\newcommand{\psib}{\overline{\psi}}
\newcommand{\sot}{\tilde{s}_0}
\newcommand{\lin}{\overline{\ell}_{\mathrm{in}}}
\newcommand{\lout}{\overline{\ell}_{\mathrm{out}}}

\newcommand{\din}{\overline{d}_{\mathrm{in}}}
\newcommand{\dout}{\overline{d}_{\mathrm{out}}}
\newcommand{\DC}{\Delta C}

\begin{document}

\title{On large deviation properties of Erd\"os-R\'enyi random graphs}

\author{Andreas Engel$^{1,2}$, R\'emi Monasson$^{2,3}$, and 
       Alexander K. Hartmann$^4$ \\
$^1$ Institut f\"ur Physik, Universit\"at Oldenburg,\\ 
26111 Oldenburg, Germany.\\
$^2$ CNRS-Laboratoire de Physique Th{\'e}orique, \\
3 rue de l'Universit\'e, 67000 Strasbourg, France.\\
$^3$ CNRS-Laboratoire de Physique Th{\'e}orique
de l'ENS,\\ 24 rue Lhomond, 75005 Paris, France.\\
$^4$ Institut f\"ur Theoretische Physik, Tammannstra{\ss}e 1,\\
 37077 G\"ottingen, Germany}

\date{\today}
\maketitle

\begin{abstract}
We show that large deviation properties of Erd\"os-R\'enyi random
graphs can be derived from the free energy of the $q$-state Potts
model of statistical mechanics. More precisely the Legendre transform
of the Potts free energy with respect to $\ln q$ is related to the
component generating function of the graph ensemble. This generalizes
the well-known mapping between typical properties of random graphs and
the $q\to 1$ limit of the Potts free energy. For exponentially rare
graphs we explicitly calculate the number of components, the size of
the giant component, the degree distributions inside and outside the
giant component, and the distribution of small
component sizes. We also perform numerical simulations which are in very
good agreement with our analytical work. Finally we demonstrate how
the same results can be derived by studying the evolution of random
graphs under the insertion of new vertices and edges, without recourse
to the thermodynamics of the Potts model.  
\end{abstract}



\section{Introduction}

Random graphs have kept being an issue of tremendous interest in
probability and graph theory ever since the seminal work by Erd\"os
and R\'enyi \cite{ErRe} more than four decades ago. In addition to fixed
edge number and fixed edge probability distributions also random
graphs with constant vertex degree \cite{Wor} or power law degree
distribution \cite{Bar,Bol2} have been investigated. 
Most of the efforts devoted to the study of the
properties of random graphs have taken advantage of the fact
that these properties undergo some concentration process in the
infinite size (number of vertices) limit. For instance, the numbers of
vertices in the largest component or the number of connected
components, which are stochastic in nature, become highly concentrated
in this limit, and with high probability do not differ from their
average values.  

For large but finite sizes, properties as the one evoked above
obviously fluctuate from graph to graph. The understanding of their
statistical deviations are important for several problems in 
statistical physics, {\em e.g.} for the life-time of metastable
states and the extremal properties of models defined on 
random graphs \cite{martin}, as well as in computer
science, {\em e.g.} for information-packet transmission in random
networks \cite{ld-queue,ld-packets}, resolution of random decision
problems with search procedures \cite{fluctuwalk,fluctudpll,fluctuvc} 
and others. Up to now apparently little attention has been paid to a
quantitative characterization of large deviations in random graph
ensembles \cite{ld-graphs,ld-graphs2}. 

The present work is intended to contribute to an improved understanding
of rare fluctuations in random graphs. Our main objective was to devise
a microscopic "mean-field" approach permitting to handle such rare
deviations in much the same way as for average properties of various
similar problems, as {\em e.g.} bootstrap and rigidity percolation
\cite{rp} and spin-glasses \cite{MPV}. The mean-field approach relies
on a statistical stability argument: a large graph is not strongly
modified when adding an edge and/or a vertex. This statement can be
translated into some self-consistent equations for the average value
of physical properties of interest, as {\em e.g.} the magnetization
for a spin system, or the probability 
of belonging to the $k$-core for bootstrap percolation. We will show in
the present work that a similar self-consistent approach can also be
successfully used to access large deviations in random graphs. 

The main property we focus on throughout this paper is the number of
connected components of a random graph. As established by Fortuin and
Kasteleyn \cite{FoKa}, several properties of random graphs with a 
{\em typical} number of components can be inferred from the
knowledge of the thermodynamics of the $q$-state Potts model on a
complete graph for values of $q$ around 1. We will show that
the thermodynamic properties for {\em general} values of $q$ can be
used to additionally characterize the properties of random graphs with
an {\em atypical} number of components. This allows us to verify the
validity of our microscopic mean-field approach. 

This paper is organized as follows. In Section 2, we introduce the basic
definitions and notations for the quantities studied. 
Section 3 is devoted to the derivation of rare graphs properties through
the study of the Potts model. We show in Section 4 how these results can
be rederived through the requirement of the statistical stability of
very large atypical graphs against the addition of a vertex and its
attached edges, or an edge. In Section 5 we describe our numerical
procedure to simulate large deviation properties of random graphs
ensembles. Some conclusion is finally proposed in Section~6.


\section{Basic notions}\label{basics}

We begin by fixing some vocabulary. For a detailed and precise
account on random graphs we refer the reader to the textbook
\cite{Bol}. A graph $G$ is a collection of {\em vertices} numbered by 
$i=1,\ldots,N$ with {\em edges} $(i,j)$, \mbox{$i\neq j$}, $i,j=1,\ldots,N$
connecting them. The number of edges is between 0 (for the empty
graph) and $N(N-1)/2$ (for the complete graph). A {\em component} of a
graph is a subset of connected vertices which are disconnect from the
rest of the graph. The {\em size} $S$ of a component is the number of
vertices it contains. Hence the empty graph consists of $N$ components
of size  1 whereas the complete graph is made from a single component
of size $N$. The {\em number} of components of a graph $G$ is denoted
by $C(G)$. We are generally interested in properties of {\em large}
graphs, $N\to\infty$.

We will consider {\em random} graphs in the sense that an edge 
between two vertices may be present or absent with a certain
probability. The various joint probabilities to be discussed below
will be denoted in the form $P(x_1,x_2,...;a_1,a_2,...)$ with the  
$x_i$ representing the random variables and the $a_i$ denoting the 
parameters of the distribution. In particular we consider random
graphs in which each pair of vertices is
connected by an edge with probability $\ga/N$ independently of all
other pairs of vertices. The parameter $\ga$ characterizes the
connectivity of the graph. Since each vertex establishes edges with
probability $\ga/N$ with all the other $N-1$ vertices $\ga$ is in the 
limit $N\to\infty$ just the typical {\em degree} of a vertex denoted
by $d^*$, giving the average number of edges emanating from it. 

More precisely, in this limit the degree $d$ of a vertex is a random
variable obeying a Poisson law with parameter $\gamma$,  
\begin{equation} \label{degdistr}
P(d;\ga) = e^{-\gamma}\; \frac{\gamma^d}{d!}.
\end{equation}
In particular, $P(d=0;\ga)=e^{-\gamma}$ is the fraction of isolated
vertices. Hence the average number of components of a random
graph of the described type is bounded
from below by $N e^{-\gamma}$. Note also that the typical degree
$d^*$ of a vertex remains finite for $N\to\infty$. Typical realization
of such random graphs are therefore {\em sparse}.  

The probability $P(G;\ga,N)$ of one particular random graph $G$ with
$N$ vertices and parameter $\ga$ derives from the binomial law, 
\begin{align}\nonumber
 P(G;\ga,N)&=\left( \frac \gamma N \right) ^{L(G)}
   \left( 1-\frac \gamma N \right)^{{N \choose 2}-L(G)}\\
    &=e^{-\frac{\ga N}{2}+\ga(\frac1 2-\frac\ga 4+\frac{L(G)} N)+ o(1)}\;
              \left( \frac \gamma N \right) ^{L(G)}, 
  \label{distriII}
\end{align}
where $L(G)=O(N)$ denotes the number of edges of graph $G$.
To describe the decomposition of a large random graph into its
components, it is convenient to introduce the probability 
$P(C;\ga,N)$ of a random graph with $N$ vertices to have $C$
components
\begin{equation}\label{probc}
P (C;\ga,N)=\sum_G P(G;\ga,N) \;\delta(C,C(G)) ,
\end{equation}
where $\delta(a,b)$ denotes the Kronecker delta.

A general observation is that for given $\ga$ and large $N$ the
probability $P(C;\ga,N)$ gets sharply peaked at some {\em typical}
value $C^*$ of $C$ and the probabilities for values of $C$ 
significantly different from $C^*$ being exponentially small in
$N$. To describe this fact more quantitatively we introduce the number
of components per vertex $c=C/N$ together with the quantity 
\begin{equation}\label{defom}
  \om(c,\ga)=\lim _{N\to \infty} \; \frac{1}{N}\ln P(C;\ga,N).
\end{equation}
Clearly $\om(c,\ga)\leq 0$ and the typical value $\ct$ of $c$ has 
$\om(\ct,\ga)=0$. Averages with $P(G;\ga,N)$ are therefore dominated
by graphs with a typical number of components.  

The focus of the present paper is on properties of random graphs which
are {\em atypical} with respect to their number of components $C$. In
order to get access to the properties of these 
graphs we introduce the {\em biased} probability distributions 
\begin{equation}\label{Pbias}
  P(G;\ga,q,N)=\frac{1}{Z(\ga,q,N)}\; P(G;\ga,N)\; q^{C(G)}, 
\end{equation}
with $Z(\ga,q,N)$ defined by
\begin{equation}\label{defZ}
  Z(\ga,q,N)=\sum_G P(G;\ga,N)\; q^{C(G)}
       =\sum_C P(C;\ga,N) \; q^C.
\end{equation}
The normalization constant $Z(\ga,q,N)$ in (\ref{Pbias}) has hence the
meaning of a component generating function of $P(G;\ga,N)$. 
Contrary to averages with $P(G;\ga,N)$ those with $P(G;\ga,q,N)$ are
dominated by graphs with an atypical number of components which is
fixed implicitly with the parameter $q$. Values of $q$
smaller than 1 shift weight to graphs with few components whereas for
$q>1$ graphs with many components dominate the distribution. The
typical case is obviously recovered for $q=1$. 

Similar to $\om(c,\ga)$ it is convenient to introduce the function 
\begin{equation}\label{defphi}
  \ph(\ga,q)=\lim _{N\to \infty} \; \frac{1}{N}\ln Z(\ga,q,N).
\end{equation}
From (\ref{defZ}) and (\ref{defom}) it follows to leading order in $N$
that 
\begin{equation}
  Z(\ga,q,N)=\int_0^1 dc \;\exp(N[\om(c,\ga)+c\ln q])
\end{equation}
and performing the integral by the Laplace method for large $N$ we
find that $\ph(\ga,q)$ and $\om(c,\ga)$ are Legendre transforms of
each other:
\begin{alignat}{3}\label{LT1}
  \ph(\ga,q)&=\max_c\,[\om(c,\ga)+c\ln q] \quad&\quad
  \om(c,\ga)&=\min_q\,[\ph(\ga,q)-c\ln q]\\\label{LT2}
        q&=\exp(-\frac{\pa \om}{\pa c})   \quad&\quad
        c&=q\frac{\pa \ph}{\pa q} 
\end{alignat}
The large deviation properties of the ensemble of random graphs as
characterized by $\om(c,\ga)$ can hence be inferred from
$\ph(\ga,q)$. In the next section we show how $\ph(\ga,q)$ can
be obtained from the statistical mechanics of the Potts model. 

For later use we also note that from differentiating (\ref{defphi})
with respect to $\ga$ we find using (\ref{defZ}) and (\ref{distriII})
to leading order in $N$
\begin{equation}\label{eqlb2}
  \lb(\ga,q)=\frac \ga 2 + \ga \frac{\pa \ph}{\pa \ga}. 
\end{equation}
Here $\lb(\ga,q)$ denotes the average number of edges per vertex in
the graph where the average is performed with the distribution
(\ref{Pbias}). 


\section{Thermodynamics of atypical graphs}\label{TD}

\subsection{The mean-field Potts model}\label{pottsmfa}
It has long been known \cite{FoKa} that certain characteristics
of random graphs are related to the thermodynamic properties of the
Potts model \cite{Pot,Wu}. The Potts model is defined in terms of an
energy function $E(\{\sigma _i\})$
depending on $N$ spin variables $\sigma_i, i=1,\ldots,N$, which
may take on $q$ distinct values  $\sigma=0,1,...,q-1$.
In the mean-field variant the energy function reads
\begin{equation}
E(\{\sigma_i\})=
    - \frac 1N\sum_{ i<j }\delta(\sigma_i,\sigma_j)
    - h \sum_{\sigma=0}^{q-1} u_\sigma \sum_i \delta(\sigma,\sigma_i),
\label{hampotts}
\end{equation}
where $h u_\sigma$ is an auxiliary field parallel to the direction
$\sigma$. 
\begin{figure}
\begin{center}
\includegraphics[width=.5\columnwidth]{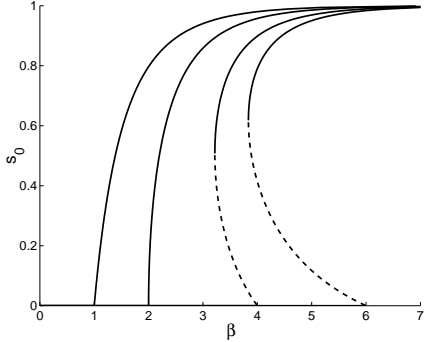} 
\caption{\label{fig:s0} Solution $\sob(\beta,q)$ of the
  saddle-point Eq.~(\ref{saddle}) as function of $\beta$ for
  $q=1,2,4,6$ (from left to right). For $q\leq 2$ the non-trivial
  solution $\sob>0$ branches off continuously from the high-temperature
  solution $\sob=0$ at $\beta=q$. For $q>2$ the new solution appears
  discontinuously at the spinodal point $\beta^{\mathrm s}<q$ by a
  subcritical bifurcation.} 
\end{center}
\end{figure}
The thermodynamic properties of the system at inverse
temperature $\beta$ can be derived from the partition function 
\begin{equation}
{\cal Z}(\beta,h,q,\{u_\sigma\},N) = \sum_{\{\sigma_i\}}
      \exp ( -\beta E(\{\sigma_i\}))
\label{partizpotts}
\end{equation}
where the sum runs over all $q^N$ spin configurations 
$\{\sigma_i\}$. A standard analysis (cf. the appendix) gives
for the free energy  
\begin{equation}\label{deffpotts}
 f(\beta,h,q,\{u_\sigma\})=-\lim_{N\to\infty} \frac 1 {\beta N} 
             \ln {\cal Z}(\beta,h,q,\{u_\sigma\},N)
\end{equation}
at $h=0$ the result
\begin{multline}\label{fpotts}
  f(\beta,q)= \mathop{\rm extr}_{s_0}
  \left[-\frac 1 {2q} - \frac{q-1}{2 q} s_0^2 
        -\frac 1 \beta \ln q\right.\\
  \left.  +\frac{1+(q-1)s_0}{\beta q}\ln(1+(q-1)s_0) +\frac{q-1}{\beta q}
               (1-s_0)\ln(1-s_0)\right].
\end{multline}
The saddle-point value $\sob(\beta,q)$ extremizing the expression
in the brackets is the stable solution of the equation 
\begin{equation}\label{saddle}
  e^{\beta\sob}=\frac{1+(q-1)\sob}{1-\sob}.
\end{equation}
Clearly $\sob=0$ is always a solution of this equation. It is,
however, unstable for large $\beta$ and another, 
non-trivial solution becomes stable which describes the spontaneous
appearance of order in the low temperature phase. Fig.~\ref{fig:s0}
displays the solutions of (\ref{saddle}) as function of $\beta$ for
different values of $q$. Note the subcriticial bifurcation in
$\sob(\beta,q)$ for $q>2$. 


\subsection{Diagrammatic expansion of the Potts model} \label{diagexp}

\begin{sloppypar}
The relation between the Potts model and the random graph ensemble
introduced in section \ref{basics} becomes apparent when considering
the high-temperature expansion of the free energy (\ref{deffpotts}) of
the Potts model. Since the Kronecker 
delta can take only the values zero or unity, the partition function
(\ref{partizpotts}) can be recast into the form \cite{FoKa}
\begin{equation}\label{zpotts0}
{\cal Z}(\beta,h,q,\{u_\sigma\},N)=\sum_{\{\sigma_i\}}\;\prod_{i<j } 
  \left[1+ w \,\delta(\sigma_i,\sigma_j) \right]\;
     e^{\,\beta\, h \sum_\sigma u_\sigma \sum_i \delta(\sigma_i,\sigma)},
\end{equation}
where 
\begin{equation}\label{defw}
w=\exp (\frac \beta N) -1=\frac \beta N +O(\frac 1 {N^2}).
\end{equation}
When expanding the product appearing in (\ref{zpotts0}) we
obtain a sum of $2^{N(N-1)/2}$ terms each of which is in one--to--one 
correspondence with a graph. The $N$ vertices of this graph represent
the Potts variables $\sigma_i$, whereas an edge $(i,j)$ stands for a
factor $w \,\delta(\sigma_i,\sigma_j)$. Performing the trace over the
configurations ${\{\sigma_i\}}$ for each term in the sum, {\em i.e.} for each
graph, separately, the Kronecker deltas constrain the Potts variables
belonging to one component of the graph to the same value. As a  
result we find the Potts partition function as a sum over graphs in
the form 
\begin{equation}\label{ZPG}
{\cal Z}(\beta,h,q,\{u_\sigma\},N) =
  \sum_{G} w^{L(G)} \prod_{n=0}^{C(G)-1}
    \Big(\sum_\sigma e^{\beta h u_\sigma S_n}\Big)
\end{equation}
where the product is over all components of the graph and $S_n$
denotes the size of the $n$-th component. We will assume that $n=0$
refers to the largest component. 

From (\ref{ZPG}) and (\ref{defw}) we find  
\begin{equation}
  {\cal Z}(\beta,h=0,q,N)=\sum_G
             \left(\frac \beta N \right)^{L(G)} q^{C(G)}. 
\end{equation}
and comparison with (\ref{defZ}) and (\ref{distriII}) yields to
leading order in $N$
\begin{equation}\label{rel1}
  {\cal Z}(\ga,h=0,q,N)=e^{\frac{\ga N} 2}  Z(\ga,q,N). 
\end{equation}
\end{sloppypar}
Correspondingly from (\ref{defphi}) and (\ref{deffpotts}) it follows
that 
\begin{equation}\label{rel2}
  f(\ga,q)=-\frac 1 2 -\frac 1 \ga \ph(\ga,q). 
\end{equation}
Eqs.~(\ref{rel1}) and (\ref{rel2}) establish the relation between the
random graph ensemble defined in section \ref{basics} and the
statistical mechanics of the Potts model sketched in section
\ref{pottsmfa}. In particular we obtain from (\ref{rel2}) and
(\ref{fpotts}) 
\begin{multline}\label{resphi}
 \ph(\ga,q)= \mathop{\rm extr}_{s_0}
  \left[\frac \ga 2\,\frac{q-1} q (s_0^2-1) + \ln q \right.\\ 
    \left.     -\frac{1+(q-1)s_0} q \ln(1+(q-1)s_0) 
               - \frac{q-1} q (1-s_0)\ln(1-s_0)\right]
\end{multline}
from which $\om(c,\ga)$ follows with the help of the Legendre
transform (\ref{LT1}), (\ref{LT2}). The equation for the saddle-point
value  $\sob(\ga,q)$ in (\ref{resphi}) is from (\ref{saddle})
\begin{equation}\label{saddle2}
  e^{\ga \sob}=\frac{1+(q-1)\sob}{1-\sob}.
\end{equation}
Differentiating (\ref{ZPG}) for $u_\sigma=\delta(\sigma,0)$ with
respect to $h$, sending first 
$N\to\infty$ and then $h\to 0$ from above, one can show that the stable
solution $\sob(\ga,q)$ of (\ref{saddle2}) is nothing but the average fraction
of vertices in the largest component $s_0=S_0/N$ in an ensemble of
random graphs with biased probability (\ref{Pbias}). Hence
the phase transition in the Potts model describing the appearance of a
spontaneous magnetization at sufficiently low temperature $1/\beta$
corresponds to a percolation transition in the random graph
ensemble giving birth to a {\em giant component} with extensively many
vertices at sufficiently large connectivity parameter $\ga$. 

We also note for later convenience that using (\ref{saddle2}) in
(\ref{resphi}) the expression for $\ph(\ga,q)$ can be rewritten as 
\begin{equation}\label{resphi2}
 \ph(\ga,q)=-\frac{\ga}{2}\;\frac{q-1}{q}\;(1+\sob^2)
       -\frac{\ga \sob}{q} + \ln (q-1+e^{\ga \sob}).
\end{equation}

It is finally useful to write (\ref{ZPG}) with $\beta$ replaced by
$\ga$ in the form 
\begin{multline}
  {\cal Z}(\ga,h,q,\{u_\sigma\},N)=
        {\cal Z}(\ga,h=0,q,N)\; e^{-\frac{\ga N} 2}\\
      \Big\langle \exp\Big(\sum_{n=0}^{C(G)-1}
      \ln(\sum_\sigma e^{\ga h u_\sigma S_n(G)})-C(G)\ln q\Big)
      \Big\rangle\ ,
\end{multline}
where the average $\langle\dots\rangle$ is with respect to the biased
probability (\ref{Pbias}). 
Singling out a possible giant component of size $S_0=Ns_0$ and grouping
together all small components of the same size we then obtain
for the free energy (\ref{deffpotts})
\begin{multline}\label{fhf0}
  f(\ga,h,q,\{u_\sigma\})=f(\ga,h=0,q) +\frac1 2 -\\
  \lim_{N\to\infty} \frac 1 {\ga N} \ln
 \Big\langle \exp(N\big[\ga h s_0(G)\;
     + \sum_S \psi(S,G)\ln\sum_\sigma e^{\ga h u_\sigma S} 
     - c\ln q\big])\Big\rangle\; .
\end{multline}
Here we have introduced the number of components of size $S$ of 
graph $G$ divided by $N$ 
\begin{equation}\label{defpsi}
  \psi(S,G)=\frac 1 N \sum_{n=1}^{C(G)-1}\delta(S,S_n(G))\ .
\end{equation}
Eq.~(\ref{fhf0}) forms a suitable starting point for the
characterization of the distribution of small components 
from the Potts free energy.


\subsection{Properties of atypical graphs}\label{tdatyp}

The connection between the Potts free energy and the component
generating function of Erd\"os-R\'enyi graphs allows to 
elucidate several large deviation properties of the random
graph ensemble. First we get for the average number of
edges per vertex from (\ref{eqlb2}) and (\ref{resphi}) 
\begin{equation}\label{lbatyp}
  \lb(\ga,q)=\frac \ga {2q}\big(1+(q-1) \sob^2(\ga,q)\big)\ . 
\end{equation}
The dependence of $\lb$ on the relative size of the giant
component $\sob(\ga,q)$ for $q\neq 1$ indicates a non-trivial internal
organization of edges in rare graphs. 

\begin{figure}
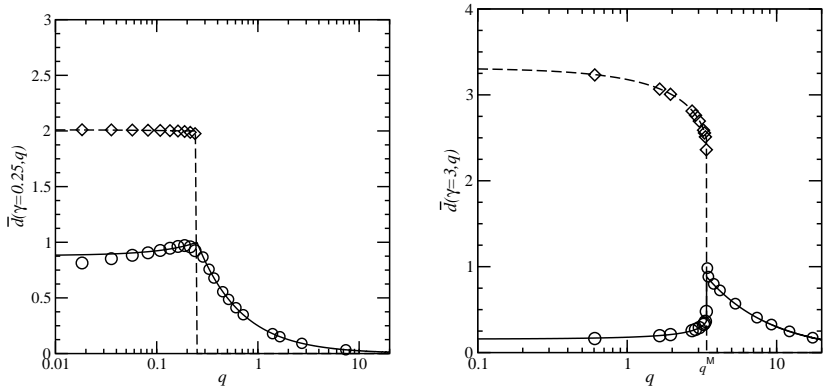

\begin{center}
\includegraphics[width=0.45\columnwidth]{d_c0.25.eps} 
\hskip .5cm
\includegraphics[width=0.45\columnwidth]{d_c3.00.eps} 
\caption{\label{fig:dc} Average degrees $\overline{d}_{\rm
    in}(\gamma,q)$ (dashed top) and $\overline{d}_{\rm
    out}(\gamma,q)$ (full) as functions of $q$ according to
  (\ref{datyp}) for $\gamma=0.25$ (left) and $\gamma=3$ (right). The
  symbols indicate numerical results (diamond=inside, circle=outside
  the giant component). The statistical error bars are much smaller
  than the symbol size. }    
\end{center}
\end{figure}

For the number of components of graphs dominating the distribution
(\ref{Pbias}) we find from (\ref{resphi}) and (\ref{LT2})
\begin{equation}\label{catyp}
  \cb(\ga,q)=\big(1-\sob(\ga,q)\big)
       \Big(1-\frac \ga {2q}\big(1-\sob(\ga,q)\big)\Big)\ .
\end{equation}
The above equations already give access to some microscopic information
on edges and vertices belonging to the giant component or to the
small components. Call $\lin$ and $\lout$ the average numbers of edges
inside and outside the giant component divided by $N$
respectively. Obviously, $\lin + \lout = \lb$. In addition, since
almost all small components are trees (cf. section \ref{sec:costs}),
the number of these components is related to the number of edges they contain
through $\cb=1-\sob - \lout$. From these two relations, we obtain
\begin{eqnarray} \label{deflinlout}
\lin (\ga,q) &=& \frac{\ga}{2q} \big( 2\, \sob + (q-2)\, s^2_0 \big)
\nonumber \\
\lout (\ga,q) &=& \frac{\ga}{2q} \big( 1- \sob \big) ^2
\end{eqnarray}
from which we deduce the average degrees 
\begin{eqnarray} \label{datyp}
\din (\ga,q) &=& \frac{\ga}{q} \big( 2 + (q-2)\, \sob \big)
\nonumber \\
\dout (\ga,q) &=& \frac{\ga}{q} \big( 1- \sob \big) 
\end{eqnarray}
of vertices inside and outside the giant component respectively. The
dependence of these degrees on $q$ for one particular value of $\ga$
is shown in Fig.~\ref{fig:dc} together with results from numerical
simulations described in section \ref{sec:numerics}.

In order to calculate the complete spectrum $\om(c,q)$ using the
Legendre transform (\ref{LT1}) we need to know $\ph(\ga,q)$ for
general real $q>0$. We have hence to study the
extremization over $s$ in (\ref{resphi}) for fixed $\ga$ and variable
$q$. This is somewhat complementary to what is done in the 
statistical mechanics of the Potts model where the free energy
(\ref{fpotts}) is minimized for integer $q\geq 2$ and different values
of $\beta$. Here we have to keep in mind that the extremum in
(\ref{resphi}) is a {\em minimum} if $q>1$ but a {\em maximum} if
$0<q<1$\footnote{The 
  reason for this is that due to the constraint (\ref{xnormal}) the
  free energy depends on $(q-1)$ variables, a number which becomes
  negative for $q<1$.}.

\begin{figure}[htb]
\begin{center}
\includegraphics[height=.45\columnwidth,angle=-90]{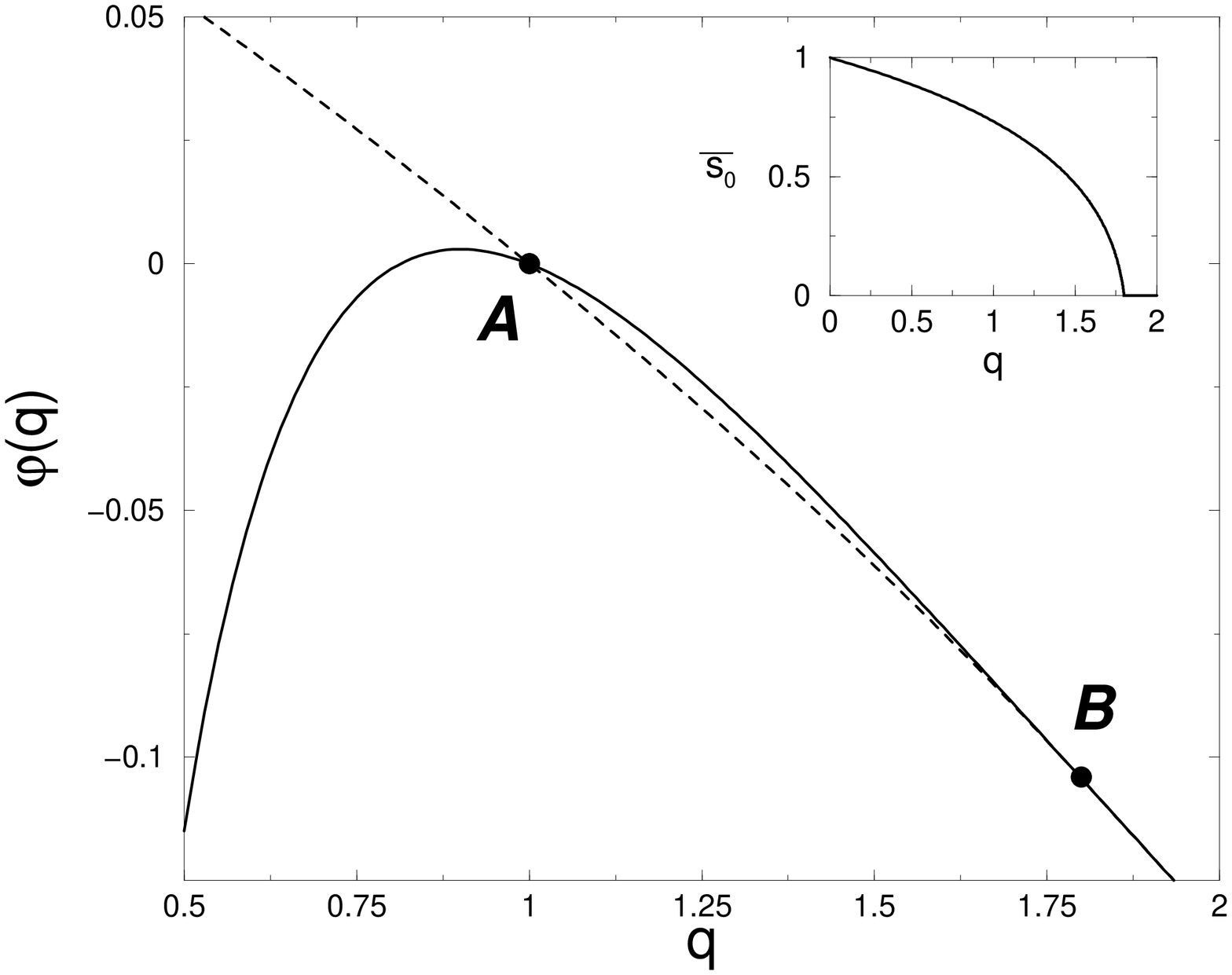} 
\hskip .5cm 
\includegraphics[height=.45\columnwidth,angle=-90]{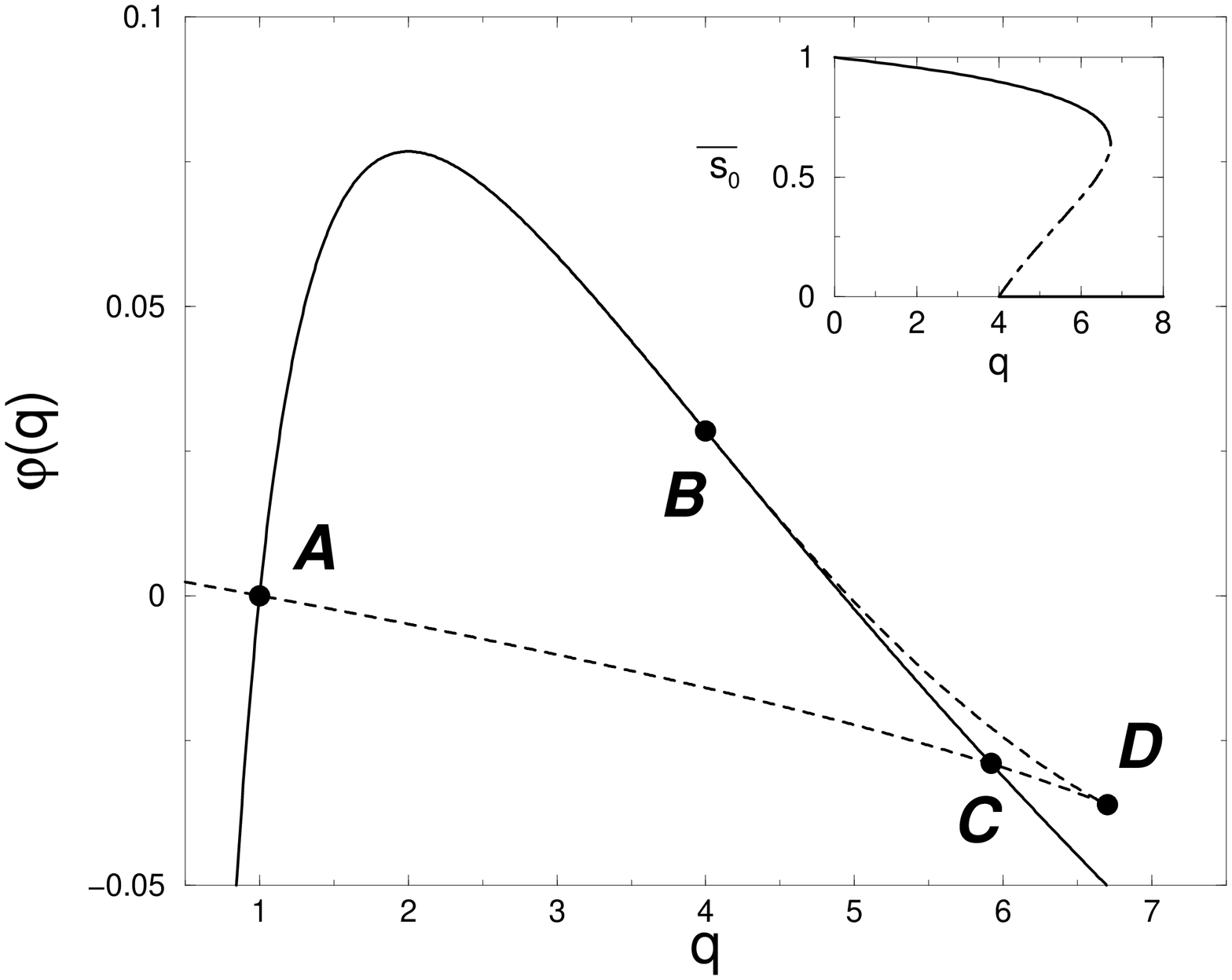} 
\caption{\label{fig:freeq} Free energy (\ref{resphi}) of the random
  graph ensemble as a function of $q$ for 
  connectivities $\ga=1.8$ (left) and $\ga =4$ (right). The full and
  dashed curve correspond to the small clusters phase ($\sob=0$) and
  the giant component phase ($\sob>0$) respectively. Both free
  energies coincide in $q=1$ (point $A$). For $\ga \leq 2$ (left), a
  second order phase transition arises when $q$ crosses $q=\ga$ (point
    $B$) with the size of the giant component approaching zero
  continuously (left inset). When $\ga>2$ (right), the transition
  takes place at point $C$ with abscissa $\qm > \ga$ and is first
  order. Both the slope of the free energy and the size of the giant
  component (right inset) are discontinuous at the
  transition. Branches $BD$ and $CD$ correspond to unstable (local
  maximum) and metastable (secondary local minimum) solutions
  respectively.}   
\end{center}
\end{figure}

The dependence of $\ph(\ga,q)$ on $q$ is  qualitatively different for
$\ga\leq 2$ and $\ga>2$ as shown in Fig.~\ref{fig:freeq}. 
For $\gamma\leq 2$ the stable solution of (\ref{saddle2}) is positive
for $q<\ga$, goes to zero for $q\to\ga$, and is identically zero for
$q>\ga$ (cf. left inset in Fig.~\ref{fig:freeq}). Accordingly
$\ph(\ga,q)$ shows a second order phase transition at $q=\ga$ as
displayed in the left part of Fig.~\ref{fig:freeq}. For $\gamma>2$ the
small-$q$ solution $\sob(\ga,q)>0$ remains stable up to $q=\qs>\ga$ and
coexists for $\ga<q<\qs$ with the solution $\sob=0$ which is stable for
$q>\ga$ as before, see right inset in Fig.~\ref{fig:freeq}. Accordingly
the phase transition is now first order and takes place at the Maxwell
point $q=\qm$ where the two values of $\ph(\ga,q)$ coincide as shown in
the right panel of Fig.~\ref{fig:freeq}. At the transition, the value
of $\sob$ jumps {\em discontinuously}, and so does the derivative of
$\ph(\ga,q)$ with respect to $q$.  

Let us now turn to the discussion of $\om(c,\ga)$. The 
bifurcation point $q=\ga$ of $\ph(\ga,q)$ maps according to
(\ref{LT1}) and (\ref{LT2}) onto $c=1/2$ for {\em all} values of
$\ga$. On the other hand the different behaviour of $\ph(\ga,q)$ for
$\ga\leq 2$ and $\ga>2$ implies qualitative differences of
$\om(c,\ga)$ in the two cases as well.  

\begin{figure}[htb]
\begin{center}
\includegraphics[height=210pt,angle=-90]{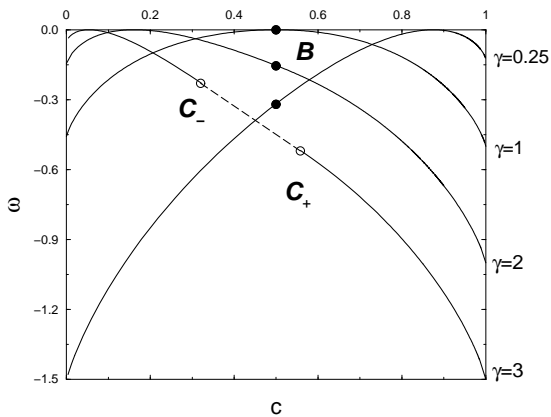} 
\caption{\label{fig:om} Logarithmic probability distribution
  $\om(c,\ga)$ of the number of components per vertex, $c$, for
  different values of the connectivity parameter $\ga=0.25, 1, 2,
  3$ (left bottom to top). $\om$ is maximal and zero for the most
  probable fraction of components $\ct$ given by (\ref{ctyp}). For
  $\ga\le 2$, there is a second order percolation transition 
  at $c=1/2$ (points B) marking the appearance of a giant component
  for $c<1/2$. When $\ga >2$, a first order transition
  separates the giant component phase (left to point $C_-$) from the
  phase without giant component (right to $C_+$). In between, both
  phases coexist and the convex hull of $\om$ is linear in $c$ (dashed
  line).} 
\end{center}
\end{figure}

For $\ga\leq 2$ we find from the Legendre transform (\ref{LT1}) that
for all values of $q$ there is exactly {\em one} corresponding value of
$c$. Accordingly the transition from the percolating phase $\sob>0$ at 
$c<1/2$ to the small component phase at $c>1/2$ is smooth as shown by
the curves for $\ga=0.25,1,2$ in Fig.~\ref{fig:om}. Except for $q=1$
the appearance of the giant component takes place in graphs with
exponentially small probabilities, $\omega(c=1/2,\ga) <0$. For $\ga<1$
this happens in the increasing part of $\om(c,\ga)$, for $\ga>1$ in
the decreasing one in accordance with the fact that the slope of 
$\om(c,\ga)$ is given by $-\ln q$, cf. (\ref{LT2}), and that $\ga=q$
at the transition. 

For $\ga>2$ the first order transition in $\ph(\ga,q)$ implies via
(\ref{LT1}) that for one particular value of $q$, namely $q=q^M$,
there are {\em two} corresponding values, $\cmone$ and $\cmtwo$, of
$c$. Hence the biased probability distribution $P(C;\ga,q^M,N)$
is bimodal and $\om(c,\ga)$ is non-convex in the interval
$\cmone<c<\cmtwo$. At the same time the Legendre transform
(\ref{LT1}) does only yield the convex hull of
$\om(c,\ga)$ and therefore includes a linear part with slope
$-\ln \qm$ interpolating between $\om(\cmone,\ga)$ and $\om(\cmtwo,\ga)$
as shown exemplarily for $\ga=3$ with the dotted line in Fig.~\ref{fig:om}. 
A random graph ensemble generated according to $P(C;\ga,q^M,N)$ is
hence inhomogeneous in the sense that it contains realizations with
$c=\cmone$ (and with giant component) and with $c=\cmtwo$ (and without
giant component). The value of $\cb$ in such an ensemble depends on
the relative fraction of these two realizations and is determined by
pre-exponential factors in $P(C;\ga,q^M,N)$. The fraction of
realizations without giant component is zero for $\cb=\cmone$, 
increases linearly with $\cb$, and reaches one at $\cb=\cmtwo$. 

\begin{figure}[tb]
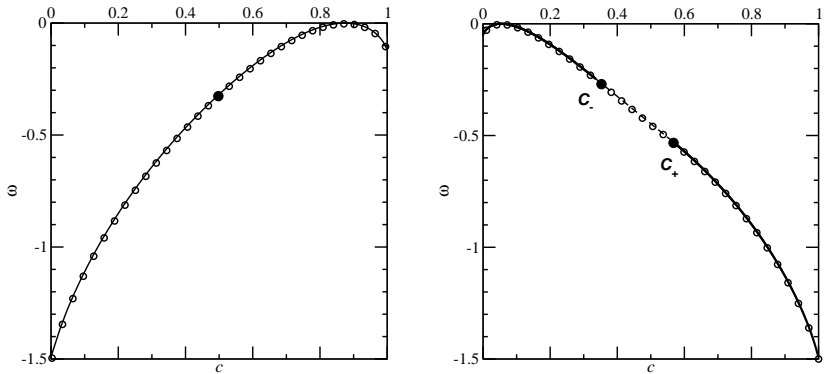

\begin{center}
\includegraphics[width=0.45\columnwidth]{omega0.25.eps} 
\hskip .5cm
\includegraphics[width=0.45\columnwidth]{omega3.eps} 
\caption{\label{fig:omnum} Comparison between analytical (full lines)
  and numerical (symbols) results for the logarithmic probability
  $\om(c,\ga)$ 
  of Erd\"os-Renyi graphs with atypical number of components for
  $\gamma=0.25$ (left) and $\ga=3$ (right). The simulations were done
  for $N=1000$ and are described in section \ref{sec:numerics}, the
  statistical error bars are much smaller than the symbol size. The
  big black dots have the same meaning as in Fig.~\ref{fig:om}.}
\end{center}
\end{figure}

The above analytical results for $\om(c,\ga)$ including the
bimodal distribution $P(C;\ga,q,N)$ for $q=q^M$ are in very good
agreement with extensive numerical simulations described in section
\ref{sec:numerics}. This is exemplified for $\ga=0.25$ and $\ga=3$ in 
Fig.~\ref{fig:omnum}.

For $c>\max(1/2,\cmtwo)$, {\em i.e.} in the region where
$\sob(\ga,q)=0$, it is possible to perform the Legendre transform
(\ref{LT1}) analytically to find 
\begin{equation}\label{oms0}
  \om(c,\ga)=-\frac \ga 2 +(1-c)(1+\ln \frac \ga 2 -\ln(1-c)).
\end{equation}
Hence we have $\om(c=1,\ga)=-\ga/2$ for all values of $\ga$ which is,
of course, consistent with Fig.~\ref{fig:om}. This result holds as
long as $\ga$ is finite. Another interesting large-$q$ limit is
obtained if $q$ and $\ga$ tend to infinity simultaneously with the
ratio $r=\ln q/\ga$ being kept constant \cite{AdA}. The tendency to
prefer graphs with many components implied by $q \to\infty$ may then
be counterbalanced by the large connectivity parameter $\ga$. In fact
for $r>1/2$ we have $q>q^M(\ga)$ and hence $\sob=0$ which brings us back
to (\ref{oms0}). On the other hand for $r\leq 1/2$ we find from 
(\ref{saddle2}) to leading order $\sob=r$ and hence from
(\ref{catyp}) $\cb=1-r$. Therefore in this case $\ga$ is large enough
to set up a giant component while the other vertices are essentially 
isolated in order to make the number of components as large as
possible.  

The opposite limit $c\to 0$ corresponds to $q\to 0$. The random graph
ensemble is for very small $q$ dominated by graphs with very few
components and for $q\to 0$ only fully connected graphs (i.e. those
with $C=1$) survive. From (\ref{saddle2}) and (\ref{resphi2}) we find
in this limit 
\begin{align}
  \sob(\ga,q)&=1-\frac{q}{e^\ga-1}+O(q^2)\\
   \ph(\ga,q)&=\ln(1-e^{-\ga})+q\;\frac{\ga(e^\ga+1)}{2(e^\ga-1)^2}+O(q^2) .
\end{align}
This results via (\ref{LT2}) and (\ref{LT1}) in 
\begin{equation}
    \cb(\ga,q)=q\;\frac{\ga(e^\ga+1)}{2(e^\ga-1)^2}+O(q^2)
\end{equation}
consistent with $C\to 1$ and 
\begin{equation}
  \om(c,\ga)=\ln(1-e^{-\ga})
       +q\;\frac{\ga(e^\ga+1)}{2(e^\ga-1)^2}(1-\ln q)+O(q^2\ln q).
\end{equation}
We hence find $\om(c=0,\ga)=\ln(1-e^{-\ga})$. This again agrees with
Fig.~\ref{fig:om} and is moreover in accordance with the known
rigorous result that the probability for an Erd\"os-R\'enyi random 
graph to be connected is asymptotically given by \mbox{$(1-e^{-\ga})^N$} 
\cite{ld-graphs}. 

We may finally extract useful information on the size distribution of
small components from the Potts free energy. Let us denote by 
\begin{equation}
  \psib(S,\ga,q)=\sum_G P(G;\ga,q,N)\; \psi(S,G)
\end{equation}
with $\psi(S,G)$ defined by (\ref{defpsi}) 
the average distribution of small components in a graph ensemble
characterized by the biased distribution $P(G;\ga,q,N)$. Consistent
with the meaning of $\psi(S,G)$ we then find to leading order in $N$
\begin{align}
  \sum_S \psib(S,\ga,q)&=\cb(\ga,q)\\
  \sum_S \psib(S,\ga,q)\; S&= 1-\sob(\ga,q)\; .
\end{align}
To get in addition an expression for the second moment of
$\psib(S,\ga,q)$ it is useful to consider the second derivative of
$f(\ga,h,q,\{u_\sigma\})$ with respect to $h$
at $h=0$ for field configurations with 
\begin{equation}\label{condu}
  u_0=1 \qquad\text{and}\qquad |u_\sigma|<1 
        \quad\text{for}\quad \sigma=1,...,q-1\;.
\end{equation}
Denoting by $\langle s_0^2\rangle_c=\langle s_0^2\rangle-\langle
s_0\rangle^2$ the second cumulant of the relative size of the giant
component and using the abbreviations
\begin{equation}
  \widehat{u}=\frac1 q \sum_\sigma u_\sigma\;\qquad\text{and}\qquad 
  \widehat{u^2}=\frac1 q \sum_\sigma u_\sigma^2
\end{equation}
on can show from (\ref{fhf0}) that 
\begin{equation}\label{fbyh1}
  -\frac1 \ga\;\frac{\pa^2 f}{\pa h^2}(\ga,h=0,q)=
  (\widehat{u^2}-\widehat{u}^2)\sum_S \psib (S,\ga,q)\; S^2      
  + (1-\widehat{u})^2\; N \langle s_0^2\rangle_c \;.
\end{equation}
On the other hand one finds from (\ref{fpotts2}) for the same quantity
after some algebra
\begin{multline}\label{fbyh2}
  -\frac1 \ga\;\frac{\pa^2 f}{\pa h^2}(\ga,h=0,q)=
   (\widehat{u^2}-\widehat{u}^2)\; \frac{q(1-\sob)}{q-\ga(1-\sob)}\\
       + (1-\widehat{u})^2\; \frac{q^2\; \sob\; (1-\sob)}
  {(q-\ga(1-\sob))(q-\ga(1-\sob)(1+(q-1)\sob))}\; .
\end{multline}

\begin{figure}[tb]
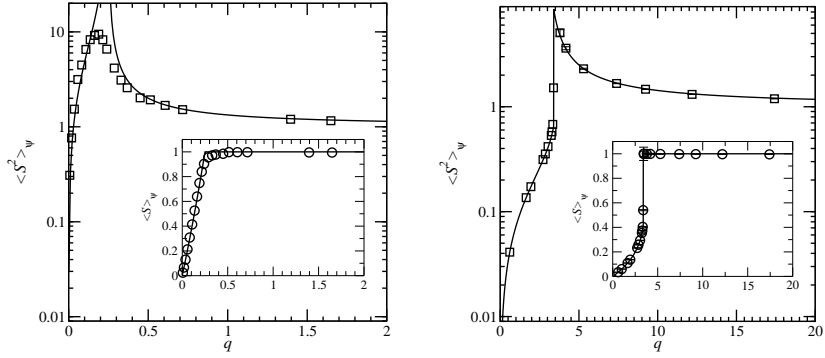

\begin{center}
\includegraphics[width=0.45\columnwidth]{psi_c0.25.eps} 
\hskip .5cm
\includegraphics[width=0.45\columnwidth]{psi_c3.00.eps} 
\caption{\label{fig:psi2} First and second moment of the distribution
  of non-extensive component sizes $\psib(S,\ga,q)$ as function of $q$
  for $\ga=0.25$ (left) and $\ga=3$ (right). Full lines are the analytical
  expressions describing the thermodynamic limit $N\to\infty$, symbols
  give results of numerical simulations for $N=1000$ described in
  section \ref{sec:numerics}.}
\end{center}
\end{figure}

Since (\ref{fbyh1}) and (\ref{fbyh2}) must be identical for any 
choice of the fields $u_\sigma$ consistent with (\ref{condu}) we are
left with 
\begin{align}\label{psiS2}
  \sum_S \psib (S,\ga,q)\; S^2
     &=\frac{q\;(1-\sob)}{q-\ga(1-\sob)}\\\label{flucs0}
  \langle s_0^2\rangle_c
     &=\frac 1 N\;\frac{q^2\; \sob\; (1-\sob)}
  {(q-\ga(1-\sob))(q-\ga(1-\sob)(1+(q-1)\sob))}\; .
\end{align}
Fig.~\ref{fig:psi2} shows the first and second moment of
$\psib(S,\ga,q)$ for two values of $\ga$ as function of $q$ together
with results from the numerical simulations. Note that for $\ga\leq 2$ the
soft transition at $\ga=q$ gives rise to a diverging second moment of 
$\psib(S,\ga,q)$ (left panel) whereas for $\ga>2$ it remains
finite at the transition (right panel). Accordingly the finite size
corrections at the transition are much larger in the first case. 

It appears to be possible to extend the above procedure to obtain also
higher moments of $\psib(S,\ga,q)$, however the calculations become
increasingly tedious. We have not been able to derive a closed expression
for the complete distribution $\psib(S,\ga,q)$ from the Potts free
energy except for the case $q=1$ which is discussed in the next section. 
A general expression for $\psib(S,\ga,q)$ will, however, be derived in
section \ref{sec:psiq} using our microscopic approach. 


\subsection{Properties of typical random graphs}\label{tdtyp}

In this subsection we rederive some of the central results for typical
graphs as special cases of our more general framework. 
As discussed in section \ref{basics} the random graph ensemble is for
large values of $N$ dominated by graphs contributing to the maximum of
$\om(c,\ga)$. Since at this maximum $\pa \om/\pa c=0$ we find from 
(\ref{LT2}) that typical properties of random graphs can be extracted
from the Potts free energy in the vicinity of $q=1$. This is well
known \cite{FoKa} and is, of course, also clear from 
the definition (\ref{Pbias}) implying \mbox{$P(G;\ga,q=1,N)=P(G;\ga,N)$}. 

\begin{figure}
\begin{center}
\includegraphics[width=100pt,angle=-90]{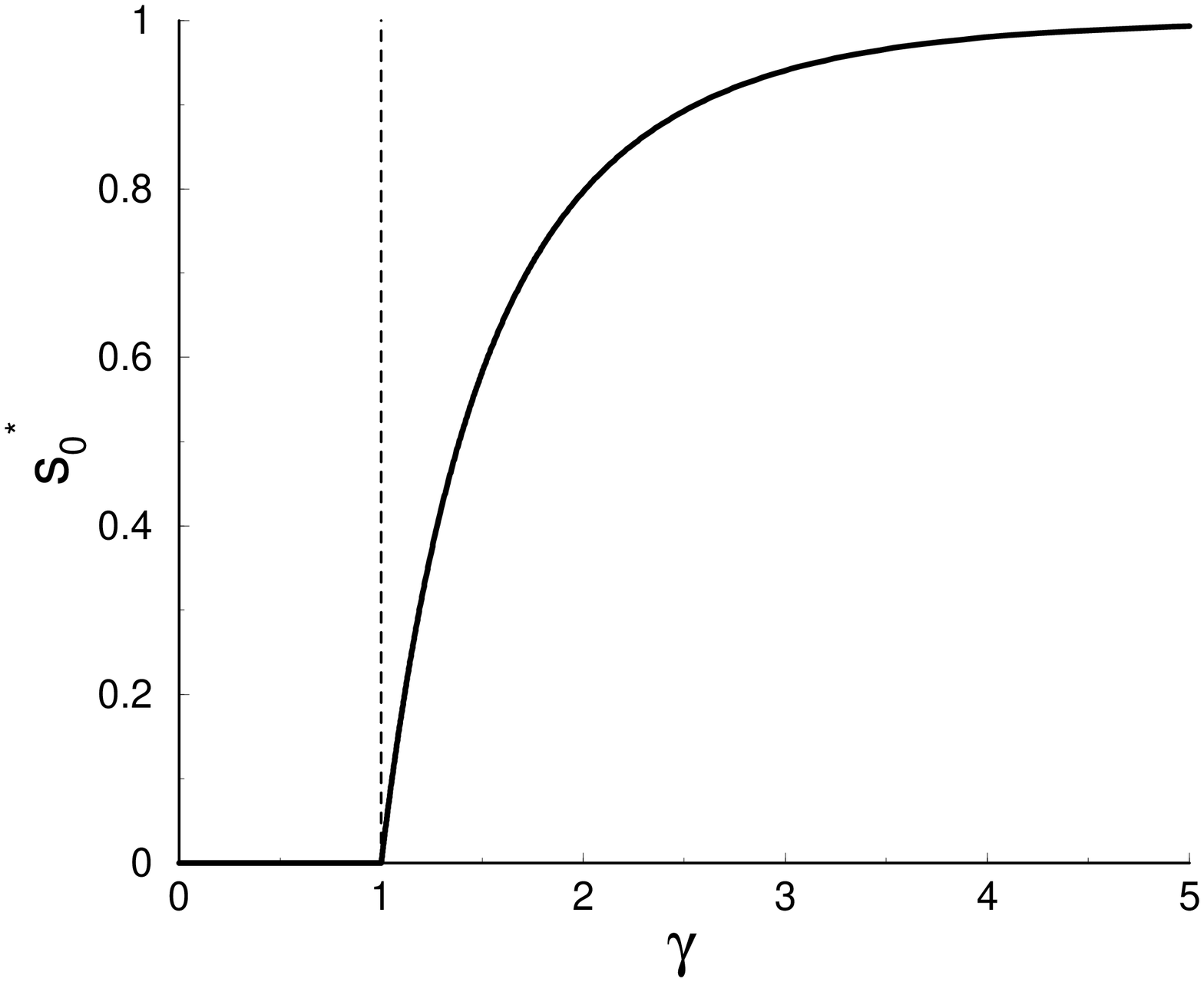} 
\hskip 1cm
\includegraphics[width=100pt,angle=-90]{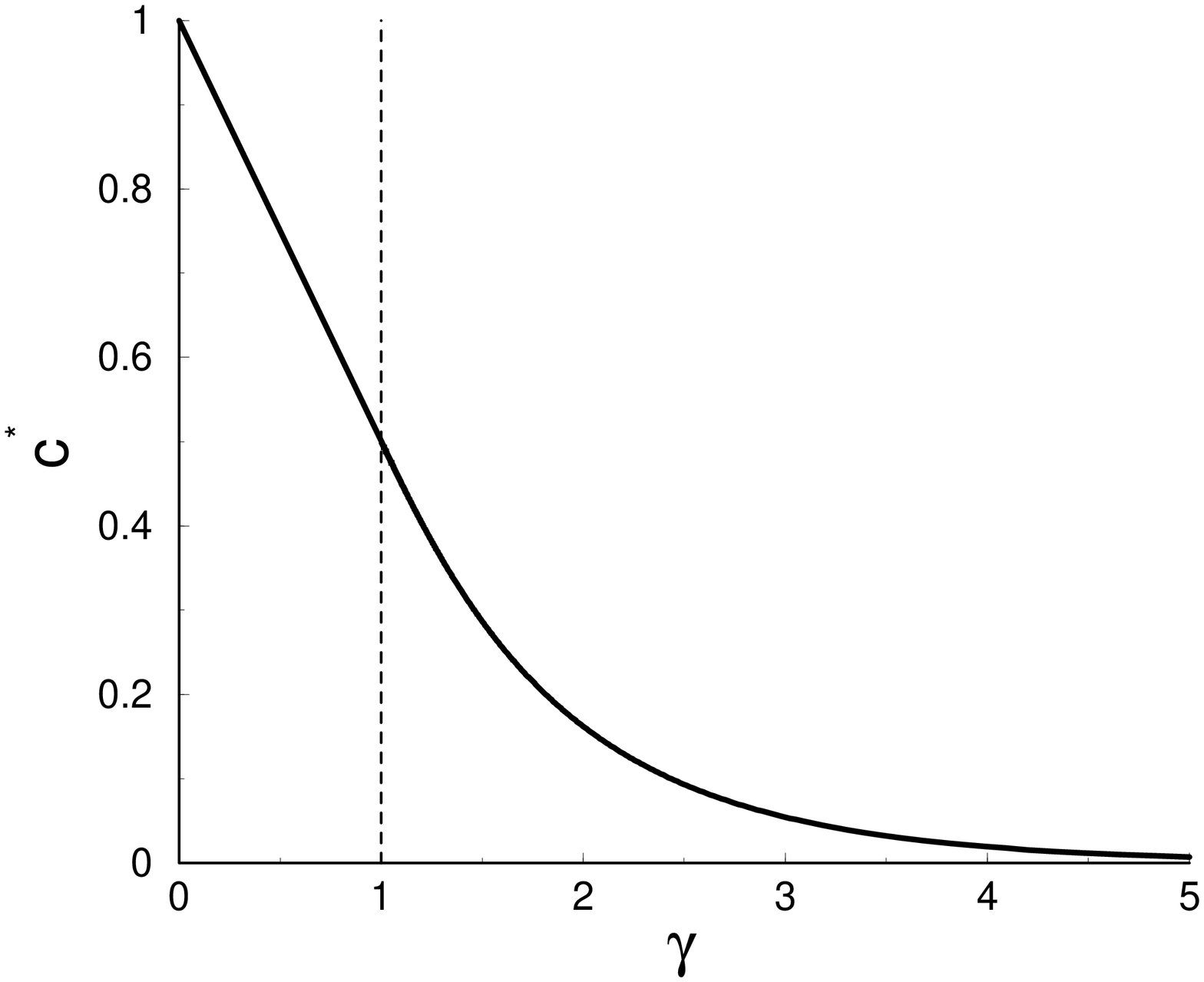} 
\caption{\label{fig:typ} Properties of typical random graphs. Shown
  are the fraction $\st$ of vertices in the largest component (left), 
and the number $\ct$ of components per vertex (right) as function of
the connectivity parameter $\ga$. The vertical dashed line, 
$\gamma=1$, indicates the location of the percolation transition.}
\end{center}
\end{figure}

Explicitly we find for the typical number of components
$\ct$ from (\ref{LT2}) and (\ref{resphi}) 
\begin{equation}\label{ctyp}
  \ct=(1-\st)(1-\frac \ga 2 (1-\st))
\end{equation}
where the relative size of the giant component $\st$ is the stable
solution of the equation 
\begin{equation}\label{styp}
  1-\st=e^{-\ga \st}
\end{equation}
which follows from (\ref{saddle2}) for $q=1$. Eqs.~(\ref{ctyp}) and
(\ref{styp}) are classical results of Erd\"os and R\'enyi \cite{ErRe}.
For small values of $\ga$ almost all components of the graph are small 
{\em trees}. Hence $\st=0$ and each edge reduces the number of
components by one. With the typical number of edges per vertex given
by (cf.~(\ref{eqlb2}) and (\ref{resphi})) 
\begin{equation}\label{lbtyp}
  \ell^*=\lb(\ga,q=1)=\frac \ga 2
\end{equation}
this implies 
$\ct=1-\ga/2$ which coincides with (\ref{ctyp}) for $s_0^*=0$. For
$\ga>1$ there is on average more than one edge 
attached to each vertex and hence the connectivity may spread out
through the whole system resulting in the emergence of a giant
component. Its size $\st$ is an increasing function of $\ga$. At the
same time the giant component has a denser connectivity than a tree
involving loops which slows down the decrease
of the number of components $\ct$ with $\ga$ as described by
(\ref{ctyp}). The dependence of $\st$ and $\ct$ on $\ga$ is shown in
Fig.~\ref{fig:typ}. The reason for the remarkable similarity between
the results (\ref{ctyp}) and (\ref{catyp}) for the number of
components in typical and atypical graphs respectively will become
clear in section \ref{sec:degreedist}. 

For the case of typical graphs considered in this subsection it is
possible to obtain some more detailed results. A simple application of
Bayes' equation for conditional probabilities \cite{Bayes} yields the 
complete degree distribution inside and outside the giant
component. The probability of vertex to have $d$ 
edges is given by (\ref{degdistr}). The probability not to belong to
the giant component conditioned to having $d$ edges is clearly
$P(\mathrm{out};d)=(1-s_0^*)^d$. The complementary probability to 
belong to the giant component is hence
$P(\mathrm{in};d)=1-(1-s_0^*)^d$. Then from Bayes' theorem we get for the
probability to have degree $d$ conditioned to being not part and being 
part of the giant component respectively
\begin{align}\label{Bayesout}
P_{\mathrm{out}}^*(d)&=\frac{P(\mathrm{out};d)\, P(d)}{P(\mathrm{out})}=
        \frac{(1-s_0^*)^d}{1-s_0^*}\, e^{-\ga}\frac{\ga^d}{d !}=
      e^{-\ga(1-s_0^*)} \frac{(\ga(1-s_0^*))^d}{d!}\\\label{Bayesin}
P_{\mathrm{in}}^*(d)&=\frac{P(\mathrm{in};d)\, P(d)}{P(\mathrm{in})}=
     \frac{1-(1-s_0^*)^d}{s_0^*}\, e^{-\ga}\frac{\ga^d}{d !}.
\end{align}
The last equality in (\ref{Bayesout}) in which we have used
(\ref{styp}) shows that the degree distribution outside the giant
component is still Poissonian. On the other hand the distribution
inside the giant component clearly deviates from a Poissonian
law. Calculating the averages of the distributions (\ref{Bayesout})
and (\ref{Bayesin}) we find 
\begin{eqnarray} \label{dtyp}
d_{\mathrm{in}}^*(\ga) &=& \ga ( 2 -\st)
\nonumber \\
d_{\mathrm{out}}^*(\ga) &=& \ga( 1- \st), 
\end{eqnarray}
consistent with (\ref{datyp}) for $q=1$. 

Also the complete distribution of component sizes can be
determined from the Potts free energy. In fact for the special choice 
$u_\sigma=\delta(\sigma,0)$ we find from (\ref{fhf0})
\begin{equation}\label{psistar1}
  \frac{\pa^2 f}{\pa q\; \pa h}(\ga,h,q=1)=
        \sum_S \psi^*(S,\ga)\;S\;e^{-\ga h S}\; ,
\end{equation}
with $\psi^*(S,\ga)=\psib(S,\ga,q=1)$. On the other hand from
(\ref{fpotts2}) we get for the same configuration of the fields
$u_\sigma$ the result 
\begin{equation}\label{psistar2}
  \frac{\pa^2 f}{\pa q\; \pa h}(\ga,h,q=1)=1-\sot(\ga,h)\; ,
\end{equation}
where $\tilde{s}_0(\ga,h)$ is the solution of 
\begin{equation}\label{hhhh}
  1-\sot=e^{-\ga(h+\sot)}.
\end{equation}
Comparing (\ref{psistar1}) and (\ref{psistar2}) we hence find 
\begin{equation} \label{h1}
\sum_{S=1}^\infty  \psi^*(S,\ga)\,S\, e^{-\ga h S} = 1 - \sot(\ga,h) \ .
\end{equation}
From (\ref{hhhh}) and (\ref{h1}) it is straightforward to produce 
equations for all the moments of $\psi^*(S,\ga)$ through successive
differentiation in $h=0$. A more direct way to obtain $\psi^*(S,\ga)$
is to get from (\ref{hhhh}) the explicit dependence of $\sot(\ga,h)$ on
$\ga$ and $h$ with the help of the Lagrange inversion theorem
\cite{Lagrange} 
\begin{equation}\label{laginv}
  1-\sot=\frac 1 \ga \sum_{S=1}^\infty 
  \frac{S^{S-1}e^{-\ga h S}}{S!} \left(\ga\,e^{-\ga}\right)^S.
\end{equation}
From (\ref{h1}) and (\ref{laginv}) we then infer 
\begin{equation}
  \sum_{S=1}^\infty  \psi^*(S,\ga)\, S\, e^{-\ga h S}=
  \frac 1 {\ga} \sum_{S=1}^\infty 
  \frac{S^{S-1}e^{-\ga h S}}{S!} \left(\ga\,e^{-\ga}\right)^S .
\end{equation}
Matching powers of $e^{-\ga h}$ we finally obtain 
\begin{equation}\label{respsityp}
\psi^*(S,\ga)=\frac 1 \ga \frac{S^{S-2}}{S!} 
           \left(\ga e^{-\ga}\right)^S,
\end{equation}
another classical result of Erd\"os and R\'enyi \cite{ErRe}. 
For the second moment of this distribution we easily find 
\begin{equation}
  \sum_{S=1}^\infty  \psi^*(S,\ga)\, S^2=
          \frac{\ga(1-\st)}{1-\ga(1-\st)}
\end{equation}
which reproduces (\ref{psiS2}) for $q=1$. We also note that
from the complementary equation (\ref{flucs0}) we get for $q=1$
\begin{equation}
  \langle s_0^2\rangle_c
     =\frac 1 N\;\frac{\st\; (1-\st)}{(q-\ga(1-\st))^2}\;,
\end{equation}
a result consistent with rigorous findings about the fluctuations of
the relative size of the giant component of typical Erd\"os-R\'enyi
graphs \cite{BBV}.


\section{Evolution of atypical graphs}\label{SM}

In the present section we will look at rare graphs from a more
microscopic point of view focusing on individual vertices and
edges. Our aim will be to rederive several of the thermodynamic 
results presented above without
reference to the Potts model. To this end we will study
the evolution of rare random graphs under the addition of a new vertex
or a new edge. This is similar in spirit to the so-called 
{\em cavity method} in the statistical mechanics of disordered systems
\cite{MPV}. The main motivation of what follows is to find an
alternative way to quantitatively characterize rare graphs. It may
be helpful in the analysis of graphs which are atypical with respect to
other properties than the number of components, as {\em e.g.} the size of
the giant component or the number of loops. In these cases the
relation to the Potts model is no longer helpful and no thermodynamic
approach seems to be known. Finally we will also derive some new
results including the complete degree distributions inside and
outside the giant component and the size distribution of non-extensive
components. These results were obtained above for typical graphs only. 


\subsection{Energetic versus entropic costs}\label{sec:costs}

From a microscopic point of view we may identify two qualitatively  
different reasons for the exponentially small probability of a
graph $G$. On the one hand the {\em number} of edges in the
graph may deviate by an extensive amount from the typical number. On
the other hand the {\em distribution} of edges among the vertices of
the graph may differ from the typical one. We will refer to these two
different sources for an exponentially small probability  as
{\em energetic} and {\em entropic} contribution respectively. 

The energetic cost is completely fixed by the probability
distribution of edges. The probability for a random graph
to have $L=\ell N$ edges is given by (cf. (\ref{distriII})) 
\begin{align}\nonumber
P(L;\ga,N) &=\binom{N^2/2}{L}\; \left(\frac{\ga}{N}\right)^{L} \;
           \left(1-\frac{\ga}{N}\right)^{N^2/2-L}\\\label{enercost}
  &=\exp\left(N\left[\ell \ln\frac{\ga}{2 \ell} 
        + \ell-\frac{\ga}{2}\right]+O(1)\right) .
\end{align}
The expression in the brackets is zero for $\ell=\ell^*=\ga/2$
reproducing (\ref{lbtyp}). It is negative for $\ell\neq \ell^*$ and
hence all other values of $\ell$ have probabilities exponentially
small in $N$. 

To leading order in $N$ we find from (\ref{enercost}) 
\begin{equation}\label{pener}
  P(L+1;\ga,N)=\frac \ga {2\,\ell} \, P(L;\ga,N),
\end{equation}
which gives the change in the energetic contribution to the
probability when one edge is added. For $\ell<\ga/2$ the
probability increases by the insertion of an edge, for  $\ell>\ga/2$
it decreases in accordance with the fact that $\ell=\ga/2$ is the
typical case. 

Let us then consider a graph with $N$ vertices, no giant component,
and an atypically large number $C$ of non-extensive components.
A possible realization of such a graph has all components as trees and
an atypically small number, $L=N-C$, of edges. However, 
this may be not the optimal way to build the graph. In fact from
(\ref{pener}) we see that the probability of the graph increases by a
factor of order 1 if we add another edge. On the other hand, in order
not to decrease at the same time the number of components we have to
put the new edge between two vertices of one of the 
{\it already existing} components. For a component with $S=O(1)$
vertices (and hence \mbox{$(S-1)$} edges) the chance to put the new
edge between two of its vertices is roughly
$(S-1)(S-2)/N^2=O(N^{-2})$. Multiplying  
by the number of components we find that the probability
not to reduce this number by putting the new edge is of
order $O(1/N)$. For large $N$ this decrease in probability cannot be
compensated by the $O(1)$ energetic gain. Hence, also in the case of
rare graphs the non-extensive components are predominantly {\it trees}. 

The situation for graphs $G$ without giant component is hence rather
clear. Since all components are trees their number is given by
$C(G)=N-L(G)$. This implies a simple relation between the
generating functions for rare graphs ($q\neq 1$) and typical graphs
($q=1$) which follows from (\ref{defZ}) and (\ref{distriII}):
\begin{equation}\label{mapgaq}
  Z_{\mathrm{s_0=0}}(\ga, q, N)=q^N Z_{\mathrm{s_0=0}}(\frac{\ga}{q},1,N)  .
\end{equation}
Therefore these graphs are characterized by an effective connectivity 
parameter $\ga/q$ and the number of edges per vertex is given by 
$\lb=\ga/(2q)$ consistent with (\ref{lbatyp}) for $\sob=0$. The
probability of such a graph is solely determined by the energetic cost
(\ref{enercost}) yielding 
\begin{equation}
P(G;\ga,q,N,\sob=0)
  =\exp\,\left(N\left[\frac{\ga}{2q}\;(1-q+\ln q)\right]\right).
\end{equation}
Replacing $q$ in this expression by $c$ according to (\ref{LT2}) and
(\ref{resphi}) we find back the logarithmic probability (\ref{oms0}).

The situation changes in the presence of a giant component of size 
$S_0=O(N)$. From the same kind of reasoning as used above it is clear,
that the entropic cost for putting an additional edge inside the giant
component is $O(1)$ and may hence well be over-compensated by the
energetic gain in probability \footnote{We expect the probability to 
  have more than one extensive component to be negligible 
  for the graphs atypical with respect to the number of components
  considered here, similarly to what happens for typical 
  random graphs \cite{unique_gc}. See also section \ref{sec:psiq}.}.  
The precise balance between energetic and entropic contributions in
this case will be investigated in the next subsection.


\subsection{Adding an edge}\label{sec:addedge}
To quantify the entropic contribution to the probability 
let us consider the probability $P(C;L)$ for a graph with $L$ edges to
have $C$ components. Upon adding one more edge the
number of components may change and we have quite generally
\begin{equation}\label{pentro}
  P(C;L+1)=\sum_{\DC} K(\DC)\, P(C+\DC;L).
\end{equation}
The kernel $K(\DC)$ is easily determined. If the new edge lies with both
ends inside the giant component of the graph the number of components
does not change at all, otherwise it is reduced by one,
\begin{equation}\label{kernel1}
  K(\DC)=s_0^2\,\delta(\DC,0) + (1-s_0^2)\,\delta(\DC,1).
\end{equation}
Combining (\ref{pener}) and (\ref{pentro}) we hence find for the
probability $P(C,L;\ga,N)=P(C;L)P(L;\ga,N)$ the evolution equation
\begin{equation}\label{addedge}
  P(C,L+1;\ga,N)=\frac \ga {2\,\ell} \,\left(
     s_0^2\, P(C,L;\ga,N)+(1-s_0^2)\, P(C+1,L;\ga,N)\right).
\end{equation}
For the biased probability 
\begin{equation}
  P(L;\ga,q,N)=\frac{1}{Z(\ga,q,N)} \sum_C  P(C,L;\ga,N)\, q^C, 
\end{equation}
where $Z(\ga,q,N)$ is defined by (\ref{defZ}) this implies
\begin{equation}\label{addedgeq}
 P(L+1;\ga,q,N)=\frac \ga {2\,\ell\, q}\,(1+(q-1)s_0^2) \,P(L;\ga,q,N)
\end{equation}
as follows from (\ref{addedge}) by multiplying with $q^C$ and summing
over $C$. Summing (\ref{addedgeq}) 
over $L$ and using the fact that this sum is dominated by graphs with 
$\ell=\lb(\ga, q)$ and $s_0=\sob(\ga,q)$ we find for the average number
of edges per vertex
\begin{equation}\label{lbatyp2}
  \lb=\frac \ga {2\,q}(1+(q-1)\sob^2).
\end{equation}
reproducing (\ref{lbatyp}).

In a similar way we may rederive the results (\ref{deflinlout}) for
the average number of edges inside and outside the giant component. To
this end we decompose the kernel (\ref{kernel1}) into contributions
corresponding to the cases of the new edge being connected to the
giant component or not 
\begin{equation}
  K(\DC)=K_{\mathrm{in}}(\DC) + K_{\mathrm{out}}(\DC).
\end{equation}
Clearly
\begin{align}
  K_{\mathrm{in}}(\DC)&=s_0^2\,\delta(\DC,0) 
             + 2s_0(1-s_0)\,\delta(\DC,1)\\
  K_{\mathrm{out}}(\DC)&=(1-s_0)^2\,\delta(\DC,1). 
\end{align}
Proceeding as above we get for the biased probability to have a graph with
$L+1$ edges with the last edge added being connected to the giant
component 
\begin{equation}
  P_{\mathrm{in}}(L+1;\ga,q,N)=
   \frac \ga {2\,\ell\, q}\,(2+(q-2)s_0^2) \,P(L;\ga,q,N).
\end{equation}
Summing over $L$ this gives 
\begin{equation}
  P_{\mathrm{in}}(\ga,q,N)=\frac \ga {2\,\lb\, q}\,(2+(q-2)\sob^2) 
\end{equation}
and hence 
\begin{equation}
  \lin=\lb \,  P_{\mathrm{in}}(\ga,q,N) =
             \frac \ga {2\, q}\,(2+(q-2)\sob^2),
\end{equation}
which is identical with (\ref{deflinlout}). Similarly one may
rederive the result for $\lout$. The results for the
average total number of edges and the average number of edges inside
and outside the giant component respectively of atypical graphs are
hence directly linked with the balance between energetic and entropic
contributions to the probability of these graphs.


\subsection{Adding a vertex}\label{addvertex}

Several interesting results may be obtained by investigating the
evolution of atypical graphs under addition of a new vertex. Compared
with the same procedure for typical graphs some special care is needed
in the present case of atypical graphs. The reason is the
following. In order to keep the statistical 
properties of the new vertex as simple as possible we will
assume that it is characterized by the simple Poissonian degree
distribution (\ref{degdistr}). In this sense we add a {\em typical}
vertex to an  {\em atypical} graph. This in turn implies a change in the
``degree of unlikeliness'' of the graph which needs to be monitored.

To make this argument more quantitative consider the following basic
step of adding one vertex. The probability of a graph $G$ with $N$
vertices and parameter $\ga$ is from (\ref{distriII}) given by 
\begin{equation}
  P(G ; \gamma , N)=
   e^{-\frac{\ga N}{2}+\ga(\frac12-\frac\ga 2+\frac{L(G)} N) +o(1)} 
           \; \left(\frac{\ga}{N}\right)^{L(G)} ,
\end{equation}
and depends on its number of vertices, $L(G)$, only.
The new vertex is assumed to have $d$ incident edges with probability 
$P(d;\ga) = e^{-\ga}\;\ga^d/d!$, cf. (\ref{degdistr}). 
There are $\binom{N}{d}$ different ways to
connect these $d$ edges with existing vertices. The new graph is
one of these possible ``wirings'', and has therefore probability
\begin{multline}\label{h1bs}
P(d;\ga)\, \frac{1}{\binom{N}{d}}\, P(G;\ga, N)\\ = 
e^{-\frac{\ga}{2}N+\ga(\frac12-\frac\ga 2+\frac{L(G)} N) -\ga+o(1)} \; 
   \left(\frac{\ga}{N}\right)^{L(G)+d} \left( 1+O(\frac 1 N) \right).
\end{multline}
But the new graph, $G'$, is one particular graph with $N+1$
vertices and $L(G')=L(G)+d$ edges. Its probability should therefore be
\begin{equation}\label{h2bs}
  P(G';\ga ,N+1)=e^{-\frac{\ga}2 (N+1)
    +\ga(\frac12-\frac\ga 2+\frac{L(G)+d}{N+1})+o(1)} \;
  \left(\frac{\ga}{N+1}\right)^{L(G)+d} . 
\end{equation}
Equality of expressions (\ref{h1bs}) and (\ref{h2bs}) imposes that 
\begin{equation}
  \frac{\ga}{2}=
        \ln   \left(\frac{N+1}{N}\right)^{L(G)+d} =
       \frac{L(G)}{N} + O(\frac1N).
\end{equation}
This is fulfilled for typical graphs since for these the
mean number of edges per vertex is indeed $\ga/2$,
cf. (\ref{lbtyp}). However, if we add a typical vertex to an 
atypical graph with $\ell\neq\ga/2$ we have to introduce an 
{\em extra multiplicative weight factor}
\begin{equation}\label{extraweight}
  w(\ga,q)=\exp\left(\frac{\ga}{2}-\ell(\ga,q)\right)
\end{equation}
in order to make the new graph an unbiased representative of the new 
ensemble. 

Let us now investigate how the probability $P(C;\ga,N)$ for a graph to 
have $C$ components as defined in (\ref{probc}) changes when we add a
new vertex. The number of components will decrease by a stochastic
amount $\DC$, and we have similarly to (\ref{pentro}) 
\begin{equation}\label{meN}
P(C;\ga,N+1) =\sum_{\DC} K(\DC) \; P(C+\DC;\ga,N) .
\end{equation}
The new kernel $K$ has now to comprise both the extra weight 
(\ref{extraweight}) of the new vertex and the probability for the change
$\DC$  when adding the new vertex. The degree $d$ of the new vertex, 
which is also the number of new edges added to the graph,
is a stochastic variable with Poissonian distribution 
$e^{-\ga}\ga^d/d!$. Of these $d$ edges, $d_0$ may be connected with 
the giant component whereas the
remaining $d-d_0$ ones are connected with small components which (with
probability 1 for $N\to\infty$) are all different from each
other. The number of components is hence reduced by $d-d_0$,  except
for the case $d_0=0$ where it changes by $d-1$. We therefore find
\begin{multline}\label{eqKN}
  K(\DC)=\sum_{d \ge 0} e^{-\frac{\ga}{2}-\ell} \;\frac{\ga^d}{d!} 
      \sum_{d_0=0}^{d}\;\binom{d}{d_0} 
       s_0^{d_0}\; (1-s_0)^{d-d_0}\;
       \delta(\DC,d-d_0-\delta(d_0,0)) 
\end{multline}
where $s_0$ is the relative size of the giant component {\em prior} to
the insertion of the new vertex.

In order to obtain results for atypical graphs from (\ref{meN}) 
we again multiply by $q^C$ and sum over $C$ to find 
\begin{equation} \label{defZr}
Z(\ga,q,N+1)=\Sigma(\ga,q)\; Z(\ga,q,N). 
\end{equation}
where  
\begin{equation}\label{defSigma}
  \Sigma(\ga,q)=\sum_{\DC} \overline{K}(\DC)\; q^{-\DC}.
\end{equation}
and $\overline{K}(\DC)$ results from $K(\DC)$ by replacing $\ell$ and
$s_0$ by $\lb(\ga,q)$ and $\sob(\ga,q)$ as  given by (\ref{saddle2})
and (\ref{lbatyp2}) respectively.
Using (\ref{eqKN}), performing the sum over $\DC$, $d_0$ and finally
over $d$ we are left with 
\begin{equation}\label{resSigma}
  \Sigma(\ga,q)=(q-1+e^{\ga \sob}) \; \exp \big( 
-\frac{\ga}{2}-\lb + \;\frac{\ga}{q}\;(1-\sob) \big) .
\end{equation}

Iterating (\ref{defZr}) we find
\begin{equation}
  \lim_{N\to\infty} \frac{1}{N} \ln Z(\ga,q,N)=
         \ln \Sigma(\ga,q).
\end{equation}
Comparison with (\ref{defphi}) and insertion of (\ref{lbatyp2}) shows
that (\ref{resSigma}) reproduces the result for the free energy
$\ph(\ga,q)$ in the form (\ref{resphi2}). To complete the rederivation
of results of the thermodynamic approach we still have to produce the
self-consistent equation (\ref{saddle2}) for $\sob(\ga,q)$. 


\subsection{The giant component} \label{secgc}

More detailed results can be obtained by again decomposing the kernel
$K(\DC)$ in (\ref{meN}) into different contributions. {\em E.g.} 
in order to reproduce the self-consistent equation for $\sob$
we decompose $K(\DC)$ into parts corresponding to the possible values
of $d_0$:  
\begin{equation}
  K(\DC)=\sum_{d_0 \ge 0} K_{d_0}(\DC),
\end{equation}
where
\begin{multline}
  K_{d_0}(\DC)=e^{\frac{\ga}{2}-\ell}\;
     \sum_{d=d_0}^{\infty}\; e^{-\ga}\;\frac{\ga^d}{d!}
     \binom{d}{d_0}s_0^{d_0}\; (1-s_0)^{d-d_0}\times  \\ 
       \delta(\DC,d-d_0-\delta(d_0,0)) .
\end{multline}
For the probability of a graph with $N+1$ vertices to have $C$
components and the last vertex added making $d_0$ connections with the
giant component we then get
\begin{equation} \label{defpm0}
  P(C,d_0;\gamma,N+1)=\sum_{\DC} K_{d_0}(\DC)\; P(C+\DC;\ga,N) .
\end{equation}

Multiplying with $q^C/Z(\ga,q,N+1)$, summing over $C$, and specifying
to the case $d_0=0$ we get for the biased probability that the new
vertex does {\em not} belong to the giant component
\begin{align}\nonumber\label{h67}
  P(d_0=0;\ga,q,N+1) &=\frac{1}{Z(\ga,q,N+1)} 
    \sum_{\DC} \overline{K}_{d_0=0}(\DC)\;q^{-\DC}\;\; Z(\ga,q,N)\\
  &=\frac{\Sigma_{d_0=0}(\ga,q)}{\Sigma(\ga,q)},
\end{align}
where we have used (\ref{defZ}), (\ref{defZr}), and introduced 
\begin{equation}\label{defSigma0}
  \Sigma_{d_0=0}(\ga,q)=\sum_{\DC} \overline{K}_{d_0=0}(\DC)\;q^{-\DC} .
\end{equation}
Performing the sum over $\DC$ and $d$ we find
\begin{equation}\label{resSigma0}
  \Sigma_{d_0=0}(\ga,q)=q\; \exp\left(-\frac \ga 2 -\lb
    +\frac{\ga}{q}(1-\sob)\right). 
\end{equation}
On the other hand for large $N$ the probability (\ref{h67}) has to be
identified with \mbox{$1-\sob(\ga,q)$}. Using (\ref{resSigma}) and
(\ref{resSigma0}) this yields finally 
\begin{equation}\label{eqsq}
  1-\sob=\frac{q}{q-1+e^{\ga \sob}}
\end{equation}
which coincides with (\ref{saddle2}). 


\subsection{The degree distribution}\label{sec:degreedist}

It is finally possible to derive expressions for the complete degree
distribution in rare graphs by decomposing the kernel $K(\DC)$ in 
(\ref{meN}) into different parts according to the value of $d$ (rather
than $d_0$ as done above)
\begin{equation}
  K(\DC)=\sum_{d \ge 0} K_{d}(\DC),
\end{equation}
where now 
\begin{multline}
  K_{d}(\DC)=e^{-\frac{\ga}{2}-\ell}\;
     \frac{\ga^d}{d!} \sum _{d_0=0}^d
     \binom{d}{d_0}s_0^{d_0}\; (1-s_0)^{d-d_0}\times  \\ 
       \delta(\DC,d-d_0-\delta(d_0,0)) .
\end{multline}
For the probability of a graph with $N+1$ vertices to have $C$
components and the last vertex added having $d$ edges this implies 
\begin{equation}\label{meN2}
P(C,d;\ga,N+1) =\sum_{\DC} K_d(\DC) \; P(C+\DC;\ga,N) ,
\end{equation}
Multiplying  by $q^C/Z(\ga,q,N+1)$ and summing over $C$ we find for 
the biased probability that the added vertex has degree $d$,
\begin{equation}
P(d;\ga,q,N+1)= \frac{\Sigma _d(\ga,q)} {\Sigma (\ga,q)}  \ ,
\end{equation}
where we have again used (\ref{defZr}), and defined
\begin{align}\label{defSigmad}
  \Sigma _d(\ga,q) &=\sum_{\DC} \overline{K}_d(\DC)\; q^{-\DC}   . 
\end{align}
Calculating explicitly the above sum, and
using (\ref{resSigma}) and (\ref{eqsq}), we obtain the degree
distribution, 
\begin{equation}\label{distridegtotq}
P(d;\ga,q) = \frac{1-\sob}q e^{-\frac{\ga}q (1-\sob)} \;
\frac{\ga^d}{d!} \; \left[ (\sob+ \;\frac{1-\sob}{q} )^d + (q-1) 
(\frac{1-\sob}{q} )^d \right] \ .
\end{equation}

This distribution reduces to the Poissonian law expected
from (\ref{degdistr}) for $q=1$ only. For $q\neq 1$ we find deviations
from a Poissonian degree distribution, where for large values of $\ga$ and
$q$ even a bimodal distribution may occur. For the average degree we obtain
from (\ref{distridegtotq}) 
\begin{equation}
\bar d(\ga,q) = \frac \ga {q}\big(1+(q-1) \sob^2(\ga,q)\big)
\end{equation}
where we have made use of the self-consistent equation (\ref{eqsq}).
Since each edge is connected with two vertices this result is
consistent with (\ref{lbatyp2}). 

While (\ref{distridegtotq}) gives the distribution of degrees for a randomly
chosen vertex in the graph, we may ask for more detailed information
depending on whether the vertex belongs, or does not belong to the giant
component. Let us call $P_{\mathrm{in}}(d;\ga,q)$ and  
$P_{\mathrm{out}}(d;\ga,q)$
the biased distributions of degrees for a vertex, respectively, inside
and outside the giant component. The generalization of the above calculation
is straightforward. $P_{\mathrm{out}}(d;\ga,q)$ and
$P_{\mathrm{in}}(d;\ga,q)$ are obtained  
from specializing the kernel $K$ to $d, d_0=0$ and $d, 1\le d_0\le d$
respectively. The calculations are very similar to the one presented
above, the results read
\begin{align}\label{Poutq}
P_{\mathrm{out}} (d;\ga,q) &=  e^{-\frac{\ga}q (1-\sob)} \;
\frac 1{d!} \; \left( \frac{\ga (1-\sob)}q \right)^d \ , \\\label{Pinq}
P_{\mathrm{in}} (d;\ga,q) &= \frac{1-\sob}{q\sob} \;
e^{-\frac{\ga}q (1-\sob)}\;
\frac{\ga^d}{d!} \; \left[ (\sob+ \;\frac{1-\sob}{q} )^d -
(\frac{1-\sob}{q} )^d \right] \ . 
\end{align}
These equations give a rather detailed description of the connectivity
in atypical graphs. For $q=1$ they reproduce (\ref{Bayesout}) and
(\ref{Bayesin}) respectively. The corresponding average degrees
$\dout$ and $\din$ are in agreement with (\ref{datyp}). The remarkable
fact that $P_{\mathrm{out}} (d;\ga,q)=P_{\mathrm{out}}^*(d,\ga/q)$
generalizes the mapping (\ref{mapgaq}) to the case $\sob\neq 0$ and
explains the similarity between the expressions (\ref{ctyp}) and
(\ref{catyp}) for the number of components in typical and atypical
graphs respectively. For large $\ga$ and $q$ the distribution
$P_{\mathrm{out}} (d;\ga,q)$ is peaked at small values of $d$ whereas 
$P_{\mathrm{in}} (d;\ga,q)$ is maximal for larger $d$ which gives rise
to the possible bimodal form of the total distribution
$P(d;\ga,q)$. We also note 
\begin{equation}
  P_{\mathrm{out}} (d=1;\ga,q)=P_{\mathrm{in}} (d=1;\ga,q)=P(d=1;\ga,q)
\end{equation}
for all values of $\ga$ and $q$ showing the special role of leaves in
the graphs. For all other values of $d$ the distributions 
$P_{\mathrm{out}} (d;\ga,q)$ and $P_{\mathrm{in}} (d;\ga,q)$ differ
from each other. Of course $P_{\mathrm{in}} (d=0;\ga,q)$ since no
isolated vertex may belong to the giant
component. 

\begin{figure}
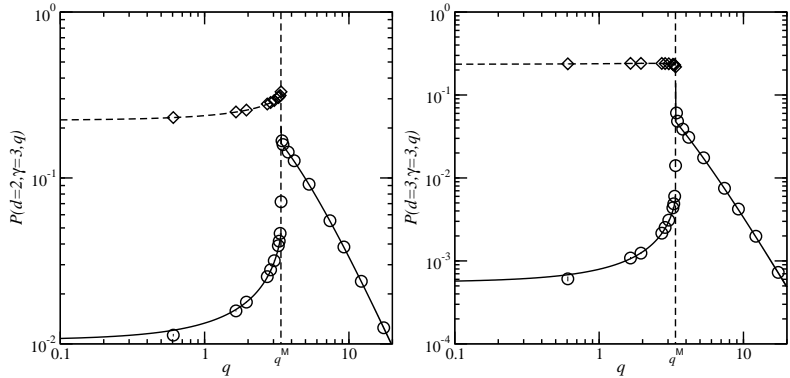

\begin{center}
\includegraphics[width=.45\columnwidth]{Pd2_c3.00.eps} 
\includegraphics[width=.45\columnwidth]{Pd3_c3.00.eps} 
\caption{\label{fig:degree} Degree distributions $P_{\rm
    in}(d;\gamma,q)$ (dashed top) and $P_{\rm out}(d;\gamma,q)$ (full)
  as given by (\ref{Poutq}) and (\ref{Pinq}) as functions of $q$ for
  $\gamma=3$, $d=2$ (left) and $d=3$ (right). 
  The numerical results are shown by symbols (diamond=inside,
  circle=outside the giant component). The dashed vertical line
  indicates the critical value $q^M$, where the giant component ceases
  to exist. The statistical error bars are much smaller than the
  symbol size. }
\end{center}
\end{figure}

Fig.~\ref{fig:degree} shows the degree distributions inside and
outside the giant component for $d=2$ and $d=3$ at $\ga=3$ as function
of $q$ together with results from numerical simulations. With
increasing $q$ the biased distribution (\ref{Pbias}) 
gets more and more dominated by graphs with many components. 
From Fig.~\ref{fig:degree} we infer that in this process both 
$P_{\mathrm{out}} (d=2;\ga,q)$ and $P_{\mathrm{in}} (d=2;\ga,q)$ 
increase. Nevertheless $P(d=2,\ga,q)$ and therefore the total
number of vertices carrying two edges decreases due to the 
shrinking of the giant component (cf. left inset in
Fig.~\ref{fig:freeq}). 

It is interesting to note that 
\begin{align}
  P'(d)&=\frac{q \sob}{1+(q-1)\sob}\; P_{\mathrm{in}} (d;\ga,q)+
   \frac{1-\sob}{1+(q-1)\sob}\; P_{\mathrm{out}} (d;\ga,q)\\ \label{Pprime}
       &=e^{-\frac{\ga}{q}(1+(q-1)\sob)} \;
         \frac 1{d!} \; \left(\frac{\ga}{q}(1+(q-1)\sob)\right)^d ,
\end{align}
{\em i.e.}, $P'(d)$ is Poissonian with parameter 
\begin{equation}
  \ga'=\frac{\ga}{q}(1+(q-1)\sob).
\end{equation}
This allows the following interpretation of (\ref{Poutq}),
(\ref{Pinq}). Let us postulate that rare graphs dominating the
distribution (\ref{Pbias}) consist of independent vertices with
effective degree distribution $P'(d)$ as given by (\ref{Pprime}) and
that the probability for a vertex to belong to the giant component is
given by $P'(\mathrm{in})=q\sob/(1+(q-1)\sob)$. Accordingly the
probability not to belong to the giant component is 
$P'(\mathrm{out})=1-P'(\mathrm{in})=(1-\sob)/(1+(q-1)\sob)$. 
Repeating then the simple Bayes argument of section \ref{tdtyp} we can
reproduce the correct results for the degree distribution inside and
outside the giant component as given by (\ref{Poutq}) and
(\ref{Pinq}). In this interpretation the shift from $P(d)$ as given by
(\ref{degdistr}) to $P'(d)$ accounts for the energetic contribution to
the probability of a rare graph whereas the replacement of
$P(\mathrm{in})=s_0^*$ by $P'(\mathrm{in})$ stands for the entropic
one.  

\subsection{The distribution of small component sizes}\label{sec:psiq}

The result (\ref{Poutq}) for the degree distribution outside the giant
component allows to calculate the complete size distribution of non-extensive
components $\psib(S,\ga,q)$. As noted in subsection \ref{sec:costs},
the small components are almost certainly trees, {\em i.e.} a component
of size $S=O(1)$ involves $S-1$ edges. From the degree distribution
(\ref{Poutq}) we get for the probability $P$ to find among $N(1-\sob)$
vertices a set of $S$ vertices that make $S-1$ connections with each
other and none with the remaining ones to leading order in $N$
\begin{align}
  P&=\binom{N(1-\sob)}{S}\; \left(\frac{\ga}{qN}\right)^{S-1}\;
      \left(1-\frac{\ga}{qN}\right)^{\big(N(1-\sob)-S\big)S}\\
   &\sim \frac N {S!} \; \frac q \ga \; \big(\frac\ga q (1-\sob)\big)^S\;
             e^{-\frac\ga q (1-\sob)S}.
\end{align}
Not all of these sets form trees however, since not all of them 
are connected. The number of (unlabeled) trees of $S$
vertices is $S^{S-2}$ \cite{wilf}. For the number of small components
of size $S$ per vertex (of the complete graph) we hence find
\begin{equation}\label{respsiq}
  \psib(S,\ga,q)=\frac 1 N \;S^{S-2}\; P=
     \frac q \ga \; \frac{S^{S-2}}{S!}\;
       \left(\frac\ga q (1-\sob)\,e^{-\frac\ga q (1-\sob)}\right)^S.
\end{equation}

\begin{figure}
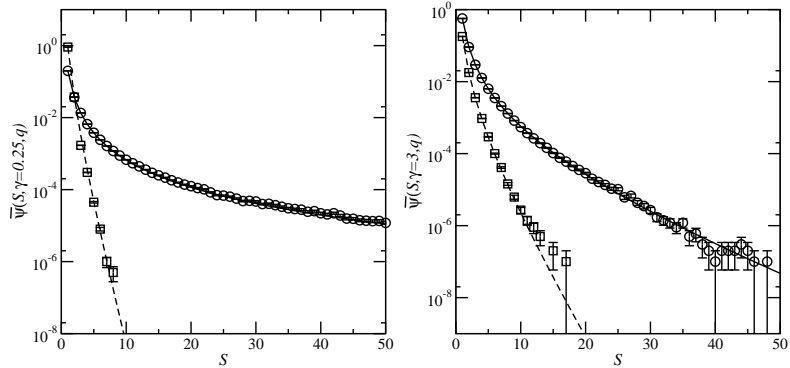

\begin{center}
\includegraphics[width=.45\columnwidth]{psib_c0.25.eps} 
\includegraphics[width=.45\columnwidth]{psib_c3.00.eps} 
\caption{\label{fig:psib} Distribution $\psib(S,\ga,q)$ of the size
  $S$ of non-extensive components in graphs with an atypical number of
  components. Left panel is for $\ga=0.25$ showing the results for
  $q=0.135$ (full line and circles) and $q=2.72$ (dashed line and
  squares) respectively. Right  panel is for $\ga=3.0$ and $q=5.29$
  (full line and circles) and $q=2.72$ (dashed line and squares)
  respectively. Lines show the analytical result (\ref{respsiq}),
  symbols represent results from numerical simulations.}
\end{center}
\end{figure}
Fig.~\ref{fig:psib} compares this expression 
for $\psib(S,\ga,q)$ with results from numerical simulations described
in section \ref{sec:numerics}. The agreement is again very good except
for relatively large components with correspondingly small probabilities,
$\psib(S,\ga,q)\lesssim 10^{-6}$, where the statistical error in the
simulation data prevents a meaningful comparison. 

For $q=1$ the result (\ref{respsiq}) for $\psib(S,\ga,q)$ reduces to
(\ref{respsityp}) after using (\ref{styp}). Moreover comparison of
(\ref{respsiq}) with (\ref{respsityp}) shows   
\begin{equation}\label{psibpsit}
  \psib(S,\ga,q)=(1-\sob(\ga,q))\; \psi^*(S,\ga')
\end{equation}
with 
\begin{equation}\label{defgap}
  \ga'=\frac{\ga}{q}(1-\sob(\ga,q)).
\end{equation}
Hence in an atypical graph of the discussed type the vertices not
belonging to the giant component can be considered to be a typical
random graph of $N'=N(1-\sob)$ vertices with effective connectivity
parameter $\ga'$. Multiplying (\ref{respsiq}) by $S$ and summing over
$S$ we find
\begin{equation}\label{eqgap}
  \ga'=\sum_S \frac{S^{S-1}}{S!} (\ga'e^-\ga')^S
\end{equation}
implying $\ga'\leq 1$ \footnote {This result may be obtained 
  independently by using (\ref{saddle2}) in (\ref{defgap}) to get 
  \mbox{$\ga'=(1-\sob)/(q\sob)\;\ln(1+q\sob/(1-\sob))\leq 1$}
for all $q>0$ and all $\sob$ with $0\leq \sob\leq 1$.}. We have
hence always $s^*(\ga')=0$. On the one hand this implies that the
outside vertices are not able to build up a giant component of their
own and therefore shows the self-consistency of our assumption that
there is only one giant component in rare random graphs of the considered
type. On the other hand it allows to easily derive results for $q\neq 1$
from the corresponding ones for $q=1, \st=0$. For the total number of
components per vertex we find, {\em e.g.}, from (\ref{psibpsit}) and 
(\ref{ctyp}) 
\begin{align}
  \cb(\ga,q)&=\sum_S \psib(S,\ga,q)
            =(1-\sob)\;\sum_S \psi^*(S,\ga')\\
            &=(1-\sob)\;\ct(\ga')
            =(1-\sob)\;\Big(1-\frac{\ga'} 2\Big)\\
            &=(1-\sob)\;\Big(1-\frac{\ga}{2q}(1-\sob)\Big)
\end{align}
reproducing (\ref{catyp}). Similarly one may rederive the expression
(\ref{psiS2}) for the second moment of $\psib(S,\ga,q)$ which,
however, follows more directly by differentiating (\ref{eqgap}) with
respect to $\ga'$. 

We finally note that the evolution equations for the various
probability distributions employed in this section are correct in the
large $N$ limit. For finite $N$, care must be paid to the fact that
the addition of an edge or a vertex slightly changes the degrees of
the vertices giving rise to $O(1/N)$ corrections. Similar corrections
occur in the application of the cavity method to spin glasses as
discussed in chapter V in \cite{MPV}. These additional terms are,
however, irrelevant in the calculations presented above. 


\section{Numerical simulations} \label{sec:numerics}

In order to check the analytical results described above, we 
have performed Monte Carlo simulations to generate graphs with
atypical numbers of components. We have performed simulations for
graph sizes between $N=50$ and $N=10000$, the results shown are for
$N=1000$, where most simulations were performed. The rare-event
algorithm \cite{align} used works in the special case here as follows:

One starts with an initial graph $G$, {\em e.g.} a typical random
graph with connectivity $\gamma$ and calculates the number of
components $C(G)$.  The simulation is performed for a given value of
$q$  \footnote{The parameter $q$ corresponds to the temperature $T$ used in
  Ref. \cite{align} via \mbox{$q=\exp(1/T)$}. The number of components
  $C$ corresponds to the energy $H$  via \mbox{$C=-{\rm sign}(T) H$.}}.
Each Monte Carlo step consists of the following steps:

\begin{sloppypar}
\begin{itemize}
\item A trial graph $G^\prime$ is generated:
\begin{itemize} 
\item Copy $G$ to $G^\prime$
\item  Select one vertex $i$ in $G^\prime$ randomly
\item Delete all edges adjacent to $i$
\item For all other vertices $j\neq i$ generate edge $(i,j)$ with probability 
      $\gamma/N$
\end{itemize}
\item Calculate number of components $C(G^\prime)$
\item Accept $G^\prime$ as new configuration $G$ with the Metropolis
      probability $\min\{1, q^{C(G^\prime)-C(G)} \}$
\end{itemize}
\end{sloppypar}

In equilibrium, this procedure generates graphs distributed according
to the probability distribution $P(G;\gamma,q,N)$ as given by
(\ref{Pbias}). Equilibration was established in the following way. Two
runs were started with two different initial configurations. One was a
typical random graph the other one for $q<1$ ($q>1$) a graph having
minimal (maximal) number of components, {\em i.e.} a fully connected
graph (a graph without edges). In the simulation the number of
components $C(t)$ was recorded as a function of Monte Carlo sweeps
(MCS) $t$. The system was considered to be equilibrated after time
$t_0$, if $C(t_0)$ agrees within the typical fluctuations for the two
starting configurations. For $N=1000$ this was the case for all values
of $q$ after $t_0=20$ MCS ($\gamma < 3$) respectively $t_0=50$ MCS 
($\gamma=3$). Hence the system equilibrates very quickly
and does not show any sign of glassy behaviour. The average value of $C$ found
in the simulation depends on the value of $q$. For values $q<1$ the average 
number of components is preferentially small, while for $q>1$ it is high.
After equilibration, we have
taken every $t_0$ MCS graphs for analysis, $10^4$ graphs for each value of
$q$. This allows to obtain various quantities, as the degree
distributions or size distributions of components, as a function of $q$.

In order to obtain numerical results for $\om(c,\ga)$ we need to
determine $P(C;\gamma,N)$ for {\em all} values $C\in[1,N]$. Since
simulations for one given value of $q$ are dominated by graphs with 
number of components close to the typical number $N\cb(\ga,q)$
corresponding to this value of $q$, simulations at various values of
$q$ have to be combined \cite{align}. To this end one records during
the simulation for each value of $q$ the biased probability
distribution  $P(C;\gamma,q,N)=\sum_G P(G;\gamma,q,N) \delta
(C,C(G))$. From (\ref{Pbias}) and (\ref{probc}) one finds for each $q$
\begin{equation}
P(C;\gamma,N)=q^{-C}\;Z(\gamma,q,N)\;P(C;\gamma,q,N).
\label{eq:transform}
\end{equation}
In order to extract from this relation $P(C;\gamma,N)$ for values of 
$C$ around $N\cb(\ga,q)\neq N\ct$, we still have to determine
$Z(\gamma,q,N)$. This in turn can be done by starting from $q=1$, 
where the value $Z(\gamma,1,N)=1$ is known. For values of $q$ 
close to $q=1$, the measured ranges of the distributions
$P(C;\gamma,1,N)$ and $P(C;\gamma,q,N)$ overlap with each other and 
$Z(\gamma,q,N)$ can be obtained from matching both distributions in
this overlapping range. This procedure can be iterated to obtain 
$Z(\gamma,q,N)$ for values of $q$ differing by an increasing amount
from the starting point $q=1$. Using (\ref{eq:transform}) the complete
distribution $P(C;\gamma,N)$ can be determined. In our simulations
using $N=1000$ and $\gamma=0.25,1,2,3$ between 22 and 27 different
values of $q$ where sufficient to obtain $P(C;\gamma,N)$ and therefore
$\om(c,\ga)$ over the full range.


\section{Conclusion}\label{conc}

In the present paper we have investigated large deviation properties
in ensembles of Erd\"os-R\'enyi random graphs. In particular we have
studied graphs that are atypical with respect to their number of
components. We have shown that several of their properties such as
their probability, the relative size of their giant component, as well
as the second moment of the distribution of component sizes can be
obtained from the Legendre transform with respect to $\ln q$ of the
mean-field free energy of the $q$-state Potts model. This generalizes
the well-known connection  between typical properties of random
graphs and the $q\to 1$ limit of the Potts free energy. Therefore this
free energy conveys also interesting information about the random
graph ensemble for values of $q\neq 1$. In particular the well known 
first order phase transition in the mean-field Potts model for
$q>2$ gives rise to a non-convex part in the logarithmic probability
of the graphs corresponding to a bimodal probability distribution
$P(C;\ga,q,N)$. 

In a second part we have rederived these results without recourse to
the Potts model by requiring the ``statistical stability'' of the
random graph ensemble under the insertion of an additional
vertex or edge. This approach is made possible by the
mere existence of the thermodynamic limit in which the
number of vertices $N$ tends to infinity. Besides reproducing the
results obtained previously we have also pointed out some subtleties
of this method when applied to exponentially rare
configurations. Moreover we obtained as additional results the 
complete degree distribution and the size distribution of
non-extensive components in atypical graphs.

Our analytical findings describing the limit in which the number $N$
of vertices of the graphs tends to infinity are in very good agreement
with numerical simulations using a rare-event algorithm for the case
$N=1000$. 

It is well known that the typical properties of Erd\"os-R\'enyi
random graphs with fixed number of edges and fixed probability of
edges are equivalent. This equivalence does, however, not carry over
to the large deviation behaviour. In fact in an ensemble with fixed
number of edges there is no energetic contribution to the probability of a
rare graph and the large deviation characteristics will be rather
different from those studied in the present paper. 

Further work to improve our understanding of the relationship
between the two processes (one more edge or one more vertex) would
be useful. It would also be interesting to see how the microscopic
approach may be generalized for the study of graphs which 
are atypical with respect to other properties than their number of
components as, {\em e.g.} their number of vertex covers
\cite{fluctuvc}, their average degree or the size of their giant 
component, where the connection to the Potts model cannot be used. 
After completion of this work we became aware of another very 
recent application of the cavity method to characterize certain 
properties of rare random graphs~\cite{Riv}.

\appendix

\section{Appendix}

In this appendix we give some intermediate results for the derivation
of the Potts free energy (\ref{fpotts}), see also \cite{Wu}. The
explicit determination of the partition function is possible since the
energy function (\ref{hampotts}) depends on the configuration of 
spins ${\{\sigma_i\}}$ solely through the fractions  
\begin{equation}\label{fracpotts}
x ({\sigma},{\{\sigma_i\}})=\frac{1}{N} \sum_{i=1}^N 
         \delta(\sigma,\sigma_i) 
\end{equation}
of variables $\sigma_i$ equal to $\sigma$. Clearly 
\begin{equation}\label{xnormal}
  \sum_\sigma x({\sigma},{\{\sigma_i\}})=1
\end{equation}
for all ${\{\sigma_i\}}$. The energy (\ref{hampotts}) may then be
rewritten as 
\begin{equation}
E({\{\sigma_i\}})= -\frac{N}{2} \sum_\sigma 
      \big( x (\sigma,{\{\sigma_i\}})\big)^2 - 
    N h \sum_\sigma u_\sigma \; x(\sigma,\{\sigma_i\})+ O(1),  
\label{Hx}
\end{equation}
and the partition function (\ref{partizpotts}) becomes to leading
order in $N$
\begin{multline}
{\cal Z}(\beta,h,q,\{u_\sigma\})=
   \sum_{\{\sigma_i\}} \exp \left(\frac{\beta N} 2 
        \sum_\sigma x(\sigma , {\{\sigma_i\}} )^2 
      +N h \sum_\sigma u_\sigma\; x(\sigma,\{\sigma_i\})\right) \nonumber \\
 = \sum_{ \{ x(\sigma)\}}
\frac{N!}{\prod_{\sigma=0}^{q-1} [N \,x(\sigma)]!} \;
 \exp \left(\frac {\beta N} 2 \sum_\sigma  x(\sigma)^2
 +\beta N h \sum_\sigma u_\sigma\; x(\sigma)\right) \nonumber \\
 =\int\limits_0^1 \prod _{\sigma=0}^{q-1} \; dx(\sigma) \;
   \exp \left(N\left[\frac \beta 2 \sum_\sigma x(\sigma)^2 
    +\beta h \sum_\sigma u_\sigma\; x(\sigma)
    -\sum_\sigma x(\sigma)\ln x(\sigma)\right]\right).
\end{multline}
The sum and the integral over $x(\sigma)$ are restricted to the
normalized subspace defined by (\ref{xnormal}). In the limit of large
$N$ the integral may be evaluated by the Laplace
method. The Potts free energy (\ref{deffpotts}) then reads
\begin{equation}\label{fpotts2}
   f(\beta,h,q,\{u_\sigma\}) =  \mathop{\rm extr}_{\{x(\sigma )\}} 
    \left[-\frac 1 2 \sum_\sigma x(\sigma)^2 
          -h \sum_\sigma u_\sigma\; x(\sigma)
          +\frac 1 \beta\sum_\sigma x(\sigma)\ln x(\sigma)\right].
\end{equation}
We will need explicit results for the free energy and its derivatives
for $h=0$ only. A suitable ansatz to perform the extremezation in
(\ref{fpotts2}) for this case is  
\begin{align}
  x(0)&=\frac 1 q (1+(q-1) s_0)\\
  x(\sigma)&=\frac 1 q (1-s_0)\qquad\qquad \qquad 
                       \text{for} \qquad\sigma\neq 0
\end{align}
with the yet undetermined parameter $s_0$. This ansatz allows for a
possible spontaneous breaking of the Potts symmetry at low temperature
and automatically fulfills the normalization (\ref{xnormal}). It gives
rise to  
\begin{multline}\label{fofs}
  f(\beta,q)= \mathop{\rm extr}_{s_0}
  \left[-\frac 1 {2q} - \frac{q-1}{2 q} s_0^2 
        -\frac 1 \beta \ln q\right.\\
  \left.  +\frac{1+(q-1)s_0}{\beta q}\ln(1+(q-1)s_0) +\frac{q-1}{\beta q}
               (1-s_0)\ln(1-s_0)\right],
\end{multline}
which is identical with (\ref{fpotts}). Differentiating the expression
in the brackets in (\ref{fofs}) with respect to $s$ we find for the
extremum value $\sob(\beta,q)$ the equation
\begin{equation}
  e^{\beta\sob}=\frac{1+(q-1)\sob}{1-\sob},
\end{equation}
reproducing (\ref{saddle}). \\[.5cm]

{\bf Acknowledgment:}
We are grateful to J.C. Angles d'Auriac, S. Cocco, O. Martin,
A. Montanari, and M. Weigt for discussion and support during the
completion of this work. Part 
the thermodynamical calculations on the Potts model were done in
cooperation with A. M. Morgante during her Master in Theoretical
Physics. A.E. acknowledges the kind hospitality at the Laboratoire de
Physique Th\'eorique at the Universit\'e Louis Pasteur in Strasbourg
where this work was done as well as financial support from the 
{\em VolkswagenStiftung}. A.K.H. obtained financial support from the
{\em VolkswagenStiftung} (Germany) within the program
``Nachwuchsgruppen an Universit\"aten'' and from the European Community
via the Human Potential Program under contract number
HPRN-CT-2002-00307 (DYGLAGEMEM), via the High-Level Scientific
Conferences (HLSC) program, and
via the Complex Systems Network of Excellence ``Exystence''.

\end{document}